\documentclass[sigconf]{acmart}
\AtBeginDocument{%
  }
\usepackage{enumitem}
\usepackage{xspace}
  \usepackage{booktabs}
  \usepackage{dblfloatfix}

\newcommand{\meridian}{\textsc{Ptolemy}\xspace}
\newcommand{\cellql}{\textsc{CellQL}\xspace}

\copyrightyear{2026}
\acmYear{2026}
\setcopyright{cc}
\setcctype{by}
\acmConference[UIST '26]{The 39th Annual ACM Symposium on User Interface Software and Technology}{November 02--05, 2026}{Detroit, MI, USA}
\acmBooktitle{The 39th Annual ACM Symposium on User Interface Software and Technology (UIST '26), November 02--05, 2026, Detroit, MI, USA}
\acmDOI{10.1145/3830398.3830537}
\acmISBN{979-8-4007-2856-3/2026/11}

\acmSubmissionID{6950}

\newcommand{\pheading}[1]{\vspace{4px}\noindent\textbf{#1}}

\usepackage{xcolor}
\DeclareRobustCommand{\rr}[1]{\textcolor{blue}{#1}}

\DeclareRobustCommand{\rr}[1]{#1}

\begin{document}

\title{\meridian: A Semantic Map of Exploratory Data Analysis}

\author{Dylan Wootton}
\email{dwootton@mit.edu}
\orcid{0000-0002-4453-6400}
\affiliation{%
  \institution{MIT CSAIL}
  \city{Cambridge}
  \state{MA}
  \country{USA}
}

\author{Denny Bromley}
\email{dbromley@salesforce.com}
\affiliation{%
  \institution{Tableau Research}
  \city{Seattle}
  \state{WA}
  \country{USA}
}

\author{Vidya Setlur}
\email{vsetlur@salesforce.com}
\affiliation{%
  \institution{Tableau Research}
  \city{Palo Alto}
  \state{CA}
  \country{USA}
}







\renewcommand{\shortauthors}{Wootton et al.}


\begin{abstract}
A central challenge in exploratory data analysis (EDA) is keeping track of what has already been examined in order to decide what to analyze next. In practice, analysts often run dozens of analyses while building an understanding of a dataset. However, most tools provide little support for maintaining an overview of this evolving process, instead exposing only a linear history of analysis steps. These tools show sequence, what came before, but not position, how a current analysis relates to the broader space of possible analyses. As a result, analysts must mentally reconstruct which parts of the space they have explored and where gaps remain, increasing the risk of redundant work or overlooked patterns. We present \meridian, a navigational interface that externalizes analysis history as a semantic map. Each analytic step is represented as a point positioned by embeddings derived from a structured description of its effective data view (e.g., columns, filters, transformations), allowing spatial distance to reflect analytic similarity. In a mixed-methods study comparing map, canvas, and tree representations, we find that maps improve global orientation and local comparison, while ordered layouts reduce decision cost. These findings surface a trade-off between orientation and actionability, and highlight design principles for supporting strategic exploration in EDA.

\end{abstract}


\begin{teaserfigure}
  \includegraphics[width=\textwidth]{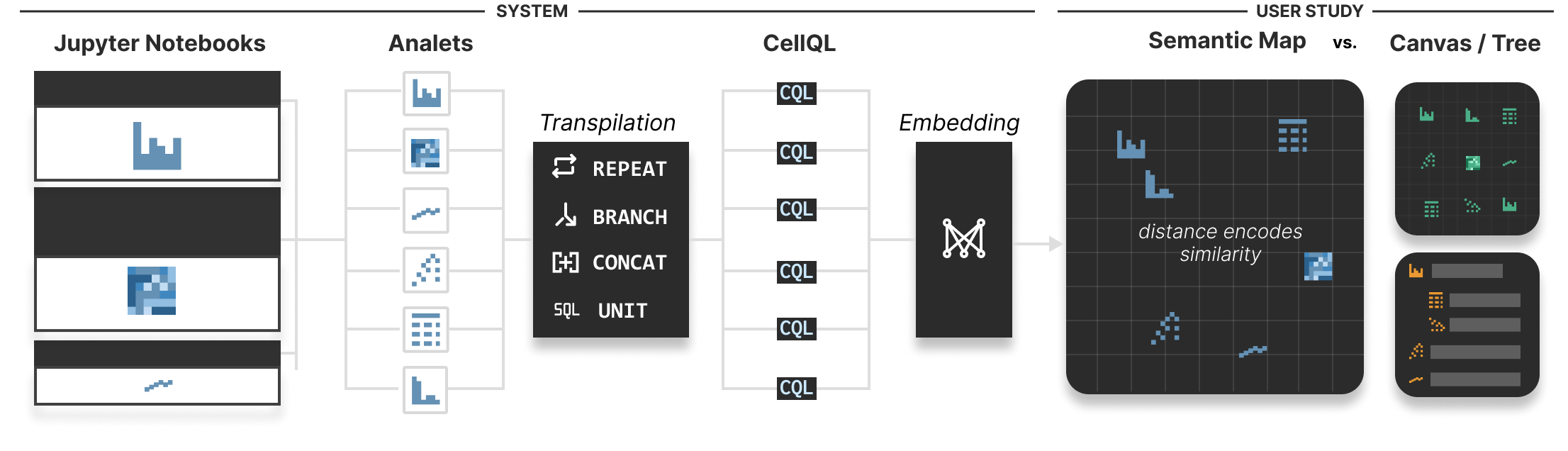}
  \caption{The \meridian~Pipeline. Python cells are extracted and turned into Analets representing a single analysis step. Analets are transpiled from Python into \cellql, a lightweight query language that models an analet's data consumption needs. From the \cellql~representation, an embedding model is used to extract vectors of semantic similarity, which are then visualized to the analyst via a UMAP. 
  Drawing on Tversky’s account of how space can express grouping, order, and distance~\cite{goos_ways_2000}, we compare a neighborhood-preserving semantic map, a semantically arbitrary grid, and an ordered hierarchy for analytical wayfinding.}
  \Description{The \meridian~Pipeline. Python cells are extracted and turned into Analets representing a single analysis step. Analets are transpiled from Python into \cellql, a lightweight query language that models an analet's data consumption needs. From the \cellql~representation, an embedding model is used to extract  vectors of semantic similarity, which are then visualized to the analyst via a UMAP. Following the distinction of \textit{ordinal}, \textit{interval}, and \textit{non-functional} spaces posed by Tversky \cite{goos_ways_2000}, we compare the affordances of trees, maps, and canvases for analytical wayfinding.} 
  \label{fig:teaser}
\end{teaserfigure}


\maketitle

\section{Introduction}
Exploratory data analysis (EDA) is iterative, branching work. Analysts generate many partial views of a dataset, decide which ones to pursue, and repeatedly return to earlier branches~\cite{battle_characterizing_2019}. In notebook workflows, however, interfaces primarily expose the current result rather than the structure of the investigation. Dataframe printouts, plots, and console outputs are effective for inspecting data, but they offer little help in understanding how an analysis is unfolding, where one is within it, or which alternative directions are nearby. As a result, analysts carry a metacognitive burden: they must mentally reconstruct their process well enough to decide whether to continue, compare, or pivot. 

These challenges are fundamentally about navigation. To navigate an ongoing analysis, analysts need to understand relationships among analysis states, not just a record of execution order. Existing notebook aids such as tables of contents~\cite{noauthor_table_2025}, code minimaps~\cite{noauthor_vscode_nodate}, and lineage visualizations~\cite{xie_waitgpt_2024,ramasamy_visualising_2023} mainly support retracing what happened before, providing what cognitive scientists call \textit{ordinal} structure. That helps answer ``\textit{what came before?}'' but not ``\textit{which earlier views are similar to this one?}'' or ``\textit{what nearby alternative should I try next?}'' This limitation makes it difficult to assess coverage: analysts cannot easily tell whether they are exploring broadly or circling the same questions. Yet in practice, analysts often bypass these tools and instead resort to data tables as proxies for reasoning about their analysis~\cite{wootton2024charting}. Tversky's distinction between \textit{ordinal} and \textit{interval} encodings~\cite{goos_ways_2000} clarifies the missing support: current tools preserve sequence, but they do not represent semantic proximity.

Many contemporary systems gesture toward interval spatialization by providing analysts a 2D canvas. Yet these canvases are usually \textit{non-functional} by default: two adjacent cells may examine unrelated aspects of the data, while near-duplicate views may be far apart. Despite appearing map-like, distances in these canvases are not grounded in a consistent metric, and users must infer structure themselves. In contrast, interval maps require a consistent unit of analysis so that notions of ``near'' and ``far'' carry semantic meaning. Importantly, we do not argue for an exhaustive map of all possible analyses, but for a sparse, semantically grounded scaffold that provides analysts with enough structure to orient, compare nearby alternatives, and make deliberate pivots.

We introduce \meridian\footnote{Named after Claudius Ptolemaeus, whose Geographia organized geographic knowledge into coordinate-based maps; our system similarly maps analysis history into a semantic coordinate space.}, a system that constructs a semantically grounded map for EDA. While existing interfaces afford \textit{ordinal} or \textit{non-functional} organization (e.g., lists, outlines, or arbitrary 2D canvases), \meridian~provides an \textit{interval} spatialization, where proximity in 2D space encodes similarity in analytic views~\cite{tversky_functional_2005}. To make those distances meaningful across heterogeneous notebook code, \meridian~employs a lightweight intermediate representation, \cellql, which extracts each cell's effective data-view signature (e.g., columns, filters, transforms, aggregations). This normalization reduces sensitivity to syntactic variation across libraries and brings semantically equivalent steps into close proximity.

Each map point represents an \textit{analet}, an analytical question roughly corresponding to a notebook cell. \meridian~embeds analets in a high-dimensional semantic space and projects them to 2D with parametric UMAP~\cite{mcinnes_umap_2020, sainburg_2009}. The interface supports interaction techniques for \textit{orientation} (locating oneself), \textit{local comparison} (inspecting nearby alternatives), \textit{diversification} (jumping to distant regions), and \textit{derivation} (creating new steps from existing ones). Our research contributions are:
\vspace*{-0.5em}
\begin{itemize}
    \item 
    \textbf{A wayfinding formulation and working system for EDA.} \meridian reifies notebook analyses as analets and organizes them into a sparse analysis space, coordinating a Semantic Map with ordinal provenance. 
    A lightweight normalization layer makes heterogeneous notebook steps comparable for local, spatial reasoning.
    \item 
    \textbf{Empirical design knowledge about spatial encodings for analytic wayfinding.} We conduct a within-subjects Comparative Structured Observation~\cite{mackay_comparative_2025} comparing a Semantic Map, a Static Canvas, and a Tree. We find the Semantic Map makes investigations easier to read as a space, but visible ordering made the next action easier to choose. These findings motivate hybrid interfaces that layer lightweight route cues onto semantically grounded maps.
\end{itemize}


\section{Related Work}
Our work is informed by research in analytical wayfinding, EDA recommendation systems, and data-fact extraction.

\subsection{Interfaces for Analytical Wayfinding}
Wayfinding research in spatial cognition provides a useful lens for evaluating how EDA interfaces support knowledge generation~\cite{siegel_development_1975,kim_acquisition_2021,miyake_navigation_2005}. This perspective distinguishes \textit{landmark}, \textit{route}, and \textit{survey} knowledge: landmarks support recognition of salient states, routes support provenance reasoning, and surveys support global reasoning about what has been covered and what remains.

\noindent \textbf{Landmark}-oriented tools provide local anchors but limited global structure. Examples include notebook tables of contents~\cite{noauthor_table_2025}, pinned views in systems such as Voyager~2~\cite{wongsuphasawat_voyager_2017}, and profiling interfaces that surface salient data properties at the current state~\cite{microsoft_power_query,lux_profiler,ydata_profiler}. These systems support local recall, but do not explicitly encode relationships among analysis steps or support reasoning about conceptual neighborhoods. 

\noindent \textbf{Route}-oriented tools emphasize lineage and sequence, including workflow/provenance systems and notebook history tools~\cite{heer2008graphical,vistrails,openlineage,rule2018janus,eckelt2024loops,cutler2020trrack,Gu_find_that_chart_26}. These systems excel at reconstructing how an analyst arrived at a result and support reproducibility. However, their representations are primarily temporal or structural. They answer ``\textit{how did I get here?}'' more readily than ``\textit{what is near here?}'' or ``\textit{what remains unexplored?}'' In open-ended EDA, this distinction becomes critical. 

\noindent \textbf{Survey}-oriented systems most directly align with our objectives. Prior work has framed EDA as hypothesis- or analysis-space exploration~\cite{suh_grammar_2023,wootton2024charting}. Systems such as Lumos and Scented View provide coverage-oriented overviews~\cite{narechania_lumos_2022,sarvghad_visualizing_2017}. However, these approaches primarily summarize coverage at the variable level, offering limited support for reasoning about operation-level similarity (e.g., differences in filters, transformations, or aggregations). 

Other systems use maps to make otherwise unbounded semantic spaces navigable. Policy Maps spatializes instances of large-language-model behavior so that practitioners can inspect coverage and author concepts within the resulting landscape~\cite{Lam_2025}. Amplio similarly treats gaps in an embedding of unstructured text as prompts for human-in-the-loop data augmentation~\cite{yeh2025exploringspaceshumanintheloopdata}. 
\meridian shares their use of spatial overview but maps executable analysis states, rather than model behaviors or data instances. 

\meridian~extends this line of work by spatializing operation-level semantics. Rather than visualizing data or variable coverage alone, we represent each analysis query through a normalized description of its effective data view and embed these representations into a similarity-based layout. This enables survey knowledge grounded not only in ``\textit{what variables were touched},'' but ``\textit{what analytic operations were performed}.''

\subsection{Recommendation Systems for EDA}
EDA recommendation systems assist analysts by suggesting charts, encodings, or statistically interesting views~\cite{mackinlay_automating_1986,wongsuphasawat_voyager_2016,wongsuphasawat_voyager_2017,demiralp_foresight_2017,vartak_seedb_2015,siddiqui_effortless_2016}. Natural language and mixed-initiative systems further infer intent from interaction context~\cite{fast_iris_2017,setlur_eviza_2016,srinivasan_orko_2018,gao_datatone_2015}. While effective for local suggestions, these systems are typically agnostic to session-level structure, lacking awareness of where the analyst has been, what is nearby, and what remains unexplored. 

\meridian~complements this literature by reframing recommendations as trajectory-level navigation over a semantic analysis space. Rather than proposing individual charts, \meridian~exposes neighborhoods of related steps and supports deliberate pivots across regions of the space. This is enabled by \cellql, which represents analysis steps in a comparable form and supports reasoning over semantic neighborhoods. 

\subsection{Data-Fact and Dashboard Generation}
Recent work in automated data-fact extraction is adjacent to our problem, as it also requires structured representations of analytic artifacts and methods for reasoning over large analytical spaces~\cite{wang_datashot_2020,vu_factflow_2025,zhao_qtsumm_2023}. These systems generate statistical statements or claims to support explanation or report generation. 

Erato~\cite{sun_erato_2022} is particularly relevant to our work: it embeds structured data facts into a vector space and interpolates between user-specified keyframes to generate intermediate facts. While both systems leverage embedding representations, the interaction goals differ. Erato supports editorial convergence, helping users author coherent linear narratives between fixed endpoints. \meridian~supports navigational divergence, helping analysts traverse an open-ended space of analysis steps, reason about semantic neighborhoods, and identify underexplored directions.

\section{System Design}


\subsection{Usage Scenario}
\label{sec:usage-scenario}

\begin{figure}[htbp]
    \centering      
    \includegraphics[width=\linewidth]{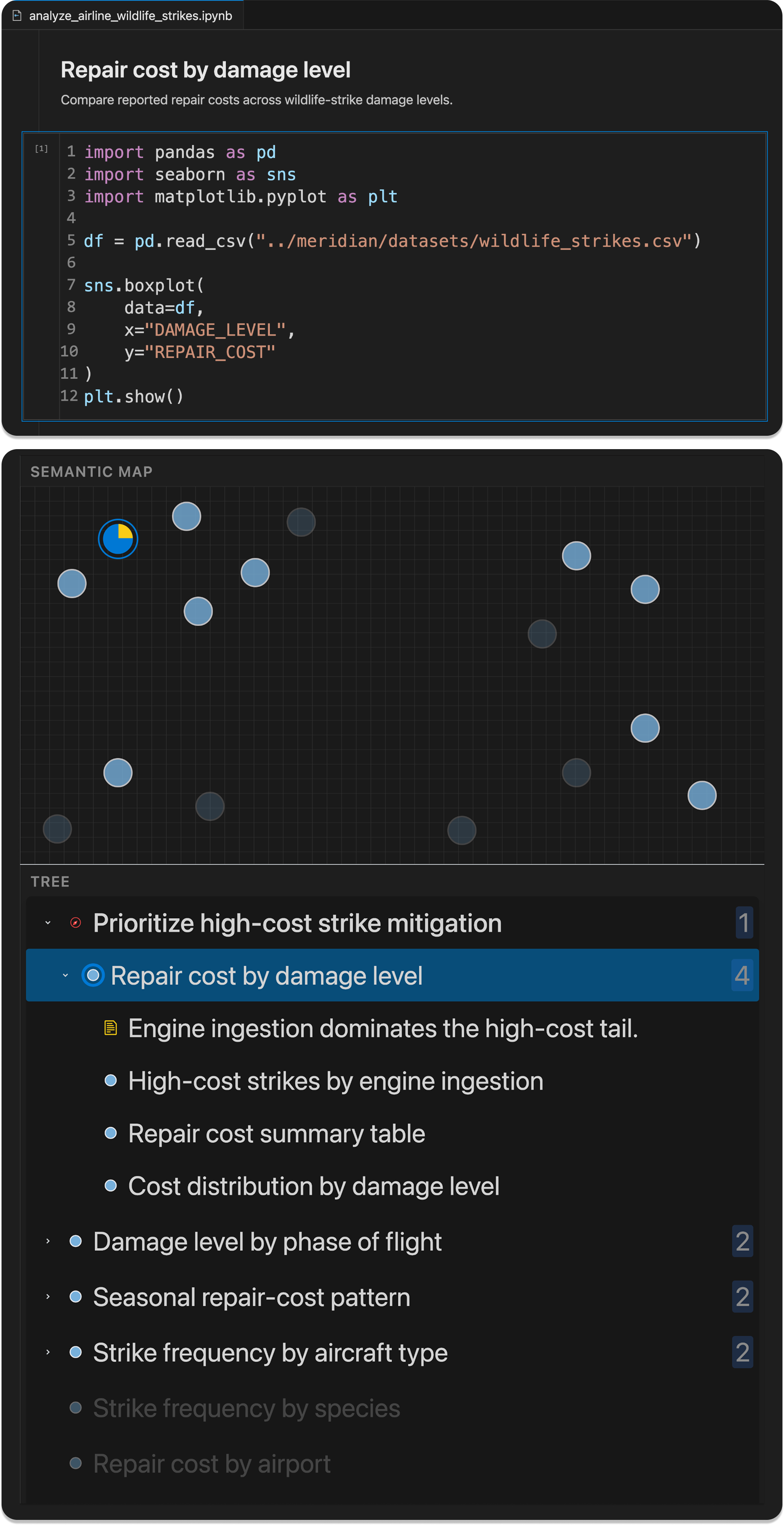}      
    \caption{          
    A usage scenario in \meridian{}. The notebook cell (top) analyzes repair cost across wildlife-strike damage levels. The Semantic Map (middle) visualizes the selected analet, shown as a selected dark blue circle among related analyses. The analet has a yellow chip to indicate a note is attached to it. The Tree (bottom) presents the same analysis session hierarchically. The selected analet is expanded to show child analyses and the recorded finding, also marked by the yellow document icon.}      
    \label{fig:user-scenario}  
\end{figure}

Maya, a data scientist at a large airline company, opens a dataset of aircraft wildlife strikes looking to determine where a fixed mitigation budget would reduce the most repair cost and operational risk.
  Each row in the dataset records one reported strike: species, aircraft type, phase of flight, altitude, a coded damage level, and repair cost.

  She loads the CSV and prints her dataframe by running \texttt{df}.
  She wants to get a sense of what she can analyze in the dataset and thus expands the \meridian~rail---a collapsible panel that sits beside the notebook.
  The map appears with one highlighted circle indicating the cell she just ran.
  This circle represents an \textit{analet}, or a self-contained analysis that can be run on the dataset, roughly corresponding to a notebook code cell.
  Nearby analets reveal other potential analyses involving repair cost, damage level, flight phase, timing, species, and aircraft type.
  Together, the neighborhood suggests several ways Maya might identify where mitigation would have the greatest effect.

  Zooming out, Maya hovers across several loose clusters.
  These are analyses generated by \meridian at the start of the session to give the space a stable shape before the analyst has authored anything.
  Hovering over an analet reveals the topic it analyzes. For example, one neighborhood concerns \texttt{repair\_cost} and \texttt{damage\_level}, another connects \texttt{phase\_of\_flight} with damage, and another counts strikes by \texttt{species}.

She clicks an analet comparing cost across damage levels and \meridian appends its code to her notebook.
She skims the code for cost distribution comparison and runs it.
The resulting distribution is severely heavy-tailed, so she follows a related analet examining high-cost strikes by engine ingestion.
This analysis shows that engine ingestions dominate the high-cost tail, and Maya records that finding in the Tree. 
Intrigued by this finding and wanting to explore more, she sets a bearing, a goal that \meridian monitors her analyses to keep her on topic. 

The Tree below places that analet inside the bearing \textit{Prioritize high-cost strike mitigation}, alongside its recorded finding and three related analyses.
Additional grouped and ungrouped analets remain visible as alternative directions Maya can pursue.

Figure~\ref{fig:user-scenario} shows this point in the analysis.
The notebook (top) contains the executable repair-cost comparison, while the \meridian~rail interface is shown below (middle and bottom).
The Semantic Map shows the selected repair-cost analet within the broader analysis space and the Tree view shows the same analet's hierarchical relationships. 

Suspecting that the tail concentrates during high-power phases of flight, she adds a filter on \texttt{phase\_of\_flight} and saves.
\meridian re-transpiles the cell and shifts its analet slightly right, now separated from its unfiltered parent.
The filtered and unfiltered views remain neighbors, making the refinement visible.

\begin{figure}[htbp]
    \centering
    \includegraphics[width=\linewidth]{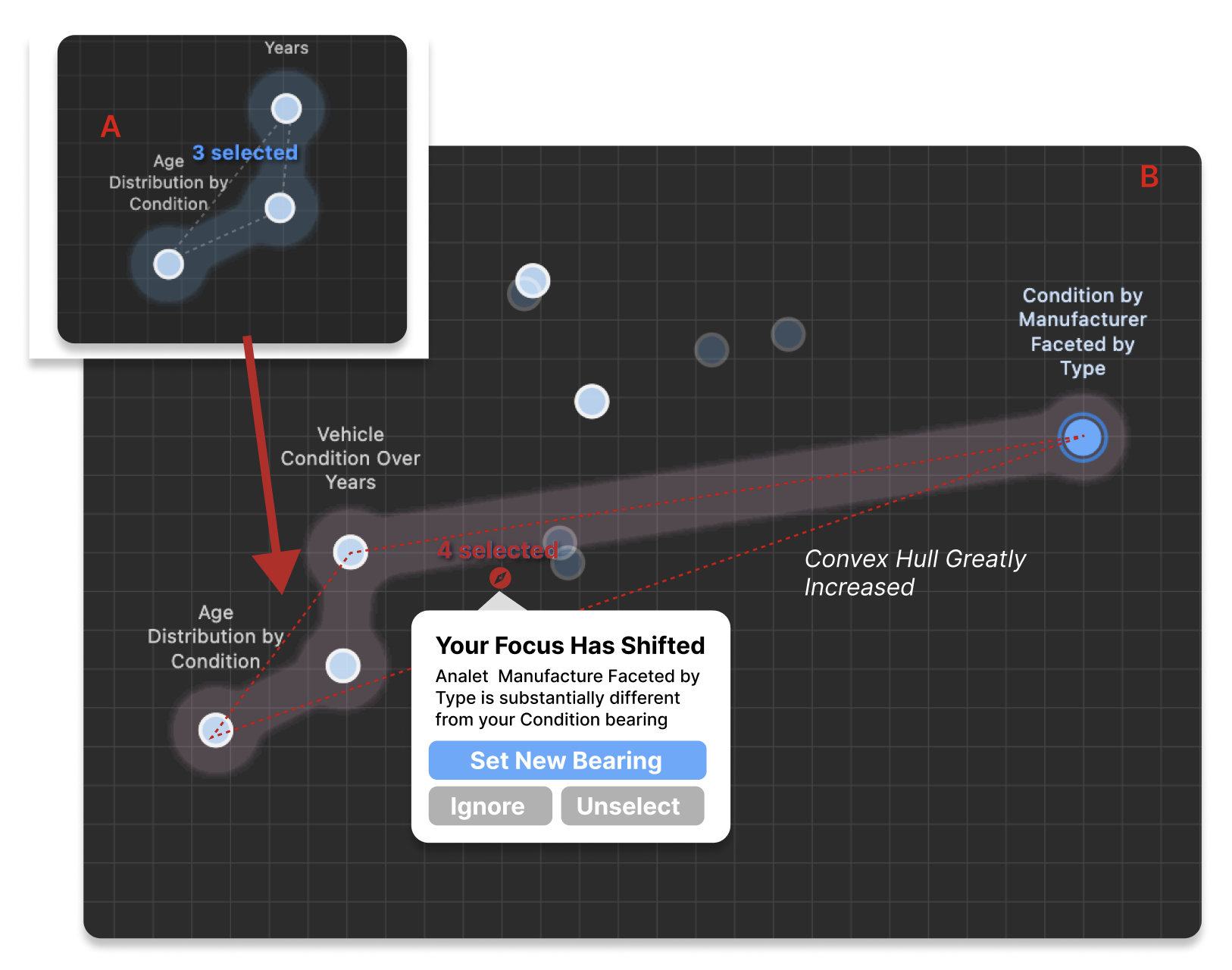}
    \caption{The analyst begins to analyze a new subset of the data prompting a \textit{semantic jump} due to the dramatic increase in the convex hull of analets under active exploration. While a bearing is activated, these jumps are monitored and surface a reflective intervention to the user.}
    \label{fig:bearings}
\end{figure}

Later on Maya develops an interest in what might be missing from her analysis. She probes the empty space in the analysis map using a \textit{fix}, which represents a request for a new analysis in a chosen region of the map. She places the fix into a sparse area, and \meridian~synthesizes a comparison of strike frequency across dawn, day, dusk, and night. 
Interested in this topic, she creates another bearing on this topic to guide her analysis. 
Curious whether the pattern holds across seasons, she continues down the timing thread by binning strikes first by month and then by hour, until a handful of unusually expensive incidents in the March bin pull her attention back to the cost column.
Each related analysis is automatically appended to this bearing's group, increasing the size of the bearing as the thread develops.
Maya then opens an unrelated analet grouping strike frequency by species.
This analet falls well outside the bearing's accumulated group, and \meridian surfaces a low-salience prompt---\textit{Your focus has shifted}---offering to set a new bearing or dismiss the jump (similar to Figure~\ref{fig:bearings}).
Maya recognizes the species question as a separate thread, dismisses the jump, and returns to the seasonal comparison; the species analet remains available for her to pick up later.

\subsection{Analysis Space and Relations}
\subsubsection{Analysis Space}
\label{subsec:analysis-space}
\meridian~models an analysis as a structured space comprising nodes and two complementary families of relations for both route and survey knowledge. Formally, we write:
\[
\textit{AnalysisSpace} := \langle \textit{Node*}, \textit{Relations} \rangle
\]
where Node represents a single analytical state (e.g., a semantically meaningful code snippet or user-authored insight) and Relations encode how these nodes connect. 

We distinguish between two relation types: \textit{Ordinal Relations} encode sequence and hierarchy, allowing users to reconstruct provenance and understand how steps unfold over time. \textit{Interval Relations} define spatial neighborhoods, enabling reasoning about similarity, continuity, and pivots in the analytic space. This split follows existing work in cognitive science on how maps encode information \cite{goos_ways_2000}; ordinal structure supports the question, \textit{``how did I get here?''}, whereas interval structure answers survey questions such as \textit{``what is nearby?''} and \textit{``where should I pivot next?''}.

Figure~\ref{fig:user-scenario} previews two views grounded in this model. The Semantic Map exposes spatial structure by positioning nodes using distances derived from the semantics of each step. The Tree exposes ordinal structure as an ordered and nested sequence.

\subsubsection{Relations}
\label{subsec:relations}
Building on the distinction between ordinal and interval encodings in cognitive science~\cite{goos_ways_2000}, \meridian~maintains two relation families:

(1) \textit{Ordinal relations} are labeled, directed links:
\[
\begin{aligned}
\textit{Ordinal} &:= \langle \textit{head}:\textit{Node}, \textit{tail}:\textit{Node}, \textit{label} \rangle,\\[4pt]
\textit{label} &\in \{\text{precedes}, \text{parent-child}, \text{group}, \ldots\}
\end{aligned}
\]

where labels include \texttt{precedes}, \texttt{parent-child}, and \texttt{group}. These support provenance reconstruction and hierarchical organization.

(2) \textit{Interval relations} define a notion of semantic proximity via a distance function over node pairs:
\[
d : \textit{Node} \times \textit{Node} \rightarrow \mathbb{R}_{\ge 0}
\]

This distance encodes semantic similarity derived from \cellql-based feature representations as discussed below.  
\meridian~surfaces ordinal and interval structure through two coordinated views, each supporting a different form of wayfinding. 

The \textit{Semantic Map} (Figure \ref{fig:user-scenario}, top) exposes interval structure by positioning nodes within a 2D semantic layout. 
Analet positions are derived from structured representations of their effective data views (via \cellql), enabling comparison across notebook code. Knowledge nodes inherit spatial placement from their associated analets or can be positioned through user interaction.
Although displayed as a 2D projection for legibility, similarity computation and recommendation logic operate over the full high-dimensional embedding space. The map serves as a perceptual scaffold for orientation. 

The \textit{Tree} (Figure, \ref{fig:user-scenario} bottom) represents the ordinal structure of the analysis as a nested hierarchy of analets and knowledge nodes. This view captures sequence, parent–child relationships, and groupings, helping users trace branching analyses. 


\subsection{Analets and CellQL}
\label{subsec:nodes}
\subsubsection{Node Types}
This subsection defines the two node types used in \meridian~and explains why they are represented within a shared space. A node is either an \textit{analet} or a \textit{knowledge node}, written
\[
\textit{Node} := \textit{Analet} \mid \textit{Knowledge}
\]

Analets represent executable analyses while knowledge nodes store user-interpreted insights or findings. A \textit{landmark} is any node that becomes cognitively salient (e.g., via annotation, pinning, or task relevance). Either node type can become a landmark, which is one reason both occupy a shared plane. 

An analet is defined as: 

\[
\textit{Analet} := \langle \textit{CellQL}, \textit{Variable\_Reference*} \rangle
\]

An analet reifies an analytic operation over data. An analet may, for example, operationalize a question such as \textit{"How does median repair cost vary by damage severity?"} or an exploratory probe such as \textit{"Show the distribution of repair costs"}. An analet may additionally carry linked representations and metadata, including a natural-language question, Python code, or description. These fields provide complementary views of the same analytic operation. In \meridian, an analet is formally defined by its CellQL representation and variable references, with linked natural-language and Python representations.

\meridian~does not position analets according to that question directly. Instead, it uses the normalized data view the question implies, so that two differently phrased questions which resolve to the same view occupy the same position in the map.

Analets function as executable landmarks in \meridian~and serve as the basis for LLM-based analysis creation. Analets enter the map from two sources, generation against the dataset schema before analysis begins (\ref{subsec:framing}) and extraction from notebook cells as the analyst works (as demonstrated in \ref{sec:usage-scenario}). In both these cases, \cellql~encodes the effective data view of that analysis and abstracts away surface-level implementation details while preserving data access and transformation semantics.

\cellql is SQL-like, but extends SQL with operators for repetition, branching, and concatenation, patterns that are common in notebook workflows but cumbersome to express in standard SQL. 
A pairplot, for example, expands into many repetitive SQL queries that obscure its shared structure with simpler views such as scatterplots, whereas \cellql~makes that structure explicit and yields compact, comparable representations of related analyses. 
Variable references maintain explicit links to the underlying data schema, enabling association with variable level metadata and tracking of variable coverage. Appendix~\ref{app:cellql-formalism} details the language design, motivates its use over alternatives such as Python, natural language descriptions, and SQL, and reports a preliminary evaluation of \cellql's robustness to syntactic variation.

As analysis proceeds, analysts accumulate conclusions that no cell encodes, such as a judgment that an apparent subgroup difference reflects collection artifact rather than real effect. A knowledge node represents this analyst-authored interpretation or insight, and is defined as: 
\[
\textit{Knowledge} := \langle \textit{Text}, \textit{Variable\_Reference*} \rangle
\]

Variable references anchor a knowledge node to the schema elements it concerns, placing it in the same reference frame as the analets that produced it.
Co-locating procedural steps and interpretive insights allows users to move fluidly between ``doing'' and ``understanding,'' supporting strategic planning rather than treating insights as detached commentary. This design aligns with prior knowledge-graph approaches for analytic reasoning~\cite{battle_what_2023,wootton2024charting}, but extends them by introducing (1) a normalized representation, \textit{\cellql}, tailored to EDA workflows and (2) interval relations for similarity-based spatial reasoning (\ref{subsec:relations}). 

\subsubsection{From Analysis Questions to \cellql}
\label{subsec:transpilation}

Embedding notebook code directly clusters analyses by syntax rather than by what they compute, separating a \textit{groupby} aggregation from an equivalent pivot despite the identical resulting view. \meridian~avoids this by normalizing through \cellql~before embedding.
To initialize the space, \meridian~uses dataset schema information, lightweight profiling statistics, any available data dictionary, and analyst task framing to generate a diverse set of candidate analysis questions in natural language (e.g., distributional summaries, subgroup comparisons, and simple relationship probes), and an LLM translates each candidate into a normalized \cellql~specification. 
These programs are embedded and projected with parametric UMAP~\cite{sainburg_2009} to form the initial spatial scaffold. During analysis, newly authored notebook steps are normalized into the same representation and inserted into the existing space using UMAP's transform operation. Parametric UMAP allows for this to be done online without refitting the entire space, so that user-authored and seeded analyses share a common reference frame. However, operations that are distinctly different from existing analets may have less accurate placements. 

\subsubsection{Framing Set}
\label{subsec:framing}
To initialize the map, \meridian~seeds each session with a curated set of 100 framing analets that serve as anchor exemplars. Without seeding, early sessions project too few analets to produce a stable layout, and each insertion would substantially rearrange the map. 
These analets are not intended to enumerate the full space of possible analyses. Instead, they provide representative coverage of common EDA patterns so that user-authored analets can be positioned within a stable reference frame. 
 We chose 100 as a pragmatic study parameter: large enough to provide useful anchor diversity, but small enough to keep latency and layout complexity manageable. Future work could examine how the size and composition of this seed set affect map stability, coverage, and recommendation quality.

\subsection{Navigation: Bearings, Fixes, and the Gazetteer}
\label{subsec:navigation}
\meridian~supports a set of interaction techniques inspired by spatial navigation metaphors to help users explore, structure, and reflect on their analytic process: \textit{Bearings} as tools for orienting, \textit{Fixes} for probing unexplored regions of the analysis space, and the \textit{Gazetteer} to accrete analyst actions into an exportable semantic layer. Bearings, Fixes, and the Gazetteer are exemplar interaction techniques made possible by representing notebook history as an embedding.

\textbf{Bearings} enable analysts to set a goal and monitor the corresponding analets that appear underneath the bearing within the analysis tree hierarchy view. Each bearing tracks the spatial footprint of its analets by computing the convex hull, i.e., the smallest region enclosing them on the map. 
As subsequent analets are added (either through creating new cells in a notebook or by manually placing them within the bearing), the bearing calculates the average area added to the hull from each added analet. 

Large expansions indicate a semantic jump (movement to a distant region), detected via a z-score over recent steps. 
When a jump occurs, \meridian~prompts reflection, aiming to make the user aware of being distracted or intervening at a natural closure of a section of their analysis. While Bearings were implemented in the full \meridian system, they were the one feature not exposed to the users during the user study due to time constraints. Thus we view the current implementation of bearings as a design probe.


\textbf{Fixes} probe underexplored regions of the map. Because distance reflects semantic similarity, empty regions may suggest untested combinations of variables, filters, or transformations. A fix converts this gap into a concrete next step. 

In ordinal space, a fix is a relative request such as ``next sibling like \textit{n}'' or ``child like \textit{n}''; the system resolves this by retrieving neighbors of \textit{n} in the embedding space and using their \cellql~as templates for generation. In interval space, a fix specifies a target region; the system retrieves the nearest framing analets and synthesizes a new \cellql~specification that adapts patterns from those neighbors. We note that sparse regions far from any seed may yield less coherent suggestions, which we mitigate through diverse initial seeding and by warning users when generating analets outside the convex hull.

Finally, the \textbf{Gazetteer} accumulates semantics over time. Notes, definitions, and derived variables are linked to their associated nodes, forming a structured, exportable semantic layer. 
This layer serves as both documentation and scaffold, enabling downstream reuse or storytelling without retracing low-level steps. Rather than requiring formal semantics upfront, \meridian~mines the semantics implicit in the work conducted by an analyst and feeds them back into subsequent sessions and downstream tools.

We implement \meridian~as a lightweight sidecar alongside the notebook, ensuring it remains unobtrusive during analysis. The sidecar collapses to a narrow rail and expands on demand; all \meridian~interactions occur within one pane. The sidecar comprises three stacked modules:

\pheading{Gazetteer.} Suggested field definitions and user-authored notes accumulate in a structured list with provenance links that trace back to the contributing analets or knowledge nodes. A footer button allows users to export this information as a semantic layer (e.g., in YAML format) for downstream use.

\pheading{Semantic Map (UMAP).} A compact 2D mini-map renders analets as points positioned via UMAP, colored by user-selected encodings (e.g., recency or variable type), and annotated with session trajectories. Hovering reveals a \cellql~summary; clicking a point scrolls the notebook to the corresponding cell. When users add new analets during a session, we position them using parametric UMAP, which projects new points into the existing embedding without recomputing the global layout. This preserves spatial stability, preventing disorienting layout shifts during interaction, at the cost of fixing the global structure to the initial framing set.

\pheading{Tree} A nested hierarchy that reflects the logical structure of the analysis, capturing parent-child relationships and serving as an organizing scaffold for tracing the lineage of ideas.

\vspace{-1.25em}

\subsection{Reading the Map: Analogical Alignment}
\label{sec:reading-the-map}

\textsc{Ptolemy} borrows a linking metaphor~\cite{zhang2026notationsevolvehistoricalanalysis} from cartography. Borrowing a map means inheriting its perceptual channels along with the meanings readers have already attached to them, and the central risk of such a borrowing is that a perceptual feature which was meaningful in the source domain arrives with a meaning that is now false. We therefore state the alignment explicitly in descending order of how trustworthy its interpretations are. 

\textbf{Local proximity} transfers cleanly. Where a cartographic map places nearby points at nearby locations, \textsc{Ptolemy} places analets with similar effective data views---overlapping columns, comparable operations---within the same neighborhood. This reading is reliable as UMAP optimizes for local neighborhood preservation, and proximity within a point's $k$ nearest neighbors reflects genuine similarity in the underlying high-dimensional space. 

\textbf{Cluster membership} follows directly. A gestalt group on the map corresponds not to a settlement but to a set of analyses over a shared family of variables and operations. Because clusters are constituted by local proximity, this reading inherits its reliability.

\textbf{Global distance} does not transfer. Two analets far apart on the map are dissimilar, but only in an ordinal sense: the reading supports ``these are unrelated'' and not ``these are twice as unrelated as those.'' UMAP retains more global structure than $t$-SNE, which is why we describe the layout as interval-like, but it offers no guarantee that would license a metric reading. Analysts should treat large separations as a weak, contextual cue.

\textbf{Empty regions} are an unreliable channel as they may occur as an artifact of projection or a genuinely unexplored area. Projection artifacts are caused by embeddings that were not present during the initial layout, and thus, we create a diverse set of seed analets in our initial framing. Each session seeds approximately 100 framing analets while only surfacing roughly 20 (Section \ref{subsec:framing}), so a region that reads as empty frequently contains seeded structure the analyst cannot see; however, we tend to find that maps stabilize around 40 analets. This also enables interaction techniques such as fixes to still draw find meaningful analyses even from empty regions. 

\textbf{Absolute position} carries no semantic content as the axes are arbitrary; however, because parametric UMAP anchors the layout to a fixed framing set, an analysis occupies the same location every time an analyst returns to it. This positional persistence enables spatial memory even where absolute position carries no meaning.

\section{Interface Evaluation}
\meridian~defines an analysis space in which analytic steps are positioned according to their semantic relationships. However, constructing this space is only part of the challenge. Analysts do not interact with embeddings or distance functions directly; rather they interact with interfaces. Whether semantic structure actually improves EDA therefore depends on how that structure is presented and navigated.

Our evaluation investigates the central question of this paper: \textit{how do different spatial encodings influence wayfinding within EDA?} To answer this question, we compare three interface conditions: map, tree, and canvas, each instantiating a different spatial logic: interval (distance-based), ordinal (sequence-based), and non-functional (layout without semantic meaning). We examine how each representation shapes orientation, survey knowledge, planning strategies, and recall of analytic history. 

\subsection{Methodology}
We conducted a within-participant study inspired by \textit{Comparative Structured Observation} (CSO)~\cite{mackay_comparative_2025}, an interventionist method that exposes participants to multiple design variants and elicits structured comparison. Each participant used three interfaces corresponding to \meridian's primary navigational representations: Semantic Map, Tree, and a Static Canvas baseline (control). 

\noindent We ground the study in two research questions: 

\noindent\textbf{RQ1:} How does a semantically grounded distance metric influence navigation during analysis? Specifically, does a semantically meaningful spatial layout (as provided by the Semantic Map) support wayfinding more effectively than a neutral grid layout, and why? 

\noindent\textbf{RQ2:} What distinct navigational affordances do the three spatial encodings (i.e., interval, ordinal, and non-functional) provide for orienting users during EDA? 

\subsubsection{Participants}
12 participants completed the study (\textit{mean} age = $29$). 
Their roles ranged from computer science PhD students to data scientists, professors, and research scientists. Participants were recruited via \textit{Upwork}, university mailing lists, and author networks, were compensated with a \$75 Amazon gift card, and self-reported at least five years of coding experience and regular use of computational notebooks. 

\subsubsection{Interfaces Tested}
Cognitive science research suggests that people construct mental representations of conceptual spaces, and that visual--spatial encodings influence what users perceive, remember, and understand~\cite{goos_ways_2000}. To examine how spatial semantics affect navigation in EDA, we compare three interface conditions as shown in Figure \ref{fig:interface-conditions}:
\begin{itemize}
  \item \textbf{Static Canvas (non-functional):} Analets are arranged in a fixed alphanumeric order on a static 2D canvas. Spatial proximity carries no semantic meaning. This condition serves as a visual baseline where layout is inert. 
  \item \textbf{Tree (ordinal):} Analets are arranged in a 1D scrollable list based on alphanumeric order. Participants organize the list into a user-defined hierarchy. This condition emphasizes route-based structure and lineage but lacks spatial embedding. 
  \item \textbf{Semantic Map (interval-like):} Analets are embedded using feature representations derived from their effective data-view structure and projected into 2D using UMAP~\cite{mcinnes_umap_2020}. Local proximity in the 2D layout reflects semantic similarity between analysis steps. Although UMAP lacks strong guarantees of global consistency, it enables intuitive neighborhood-based reasoning and weak global arrangement. We refer to this layout as ``interval-like'' due to its emphasis on relative rather than absolute distances.
\end{itemize}


Across all conditions, participants were presented with the same initial layout. Comparing results between the Map and Canvas conditions provides evidence of the effect of semantic similarity as a spatial encoding, since both use a 2D canvas but only the Map has meaningful spatial relationships. To preserve this comparison, the Canvas condition was kept fixed. Allowing participants to rearrange items would introduce user-generated structure, conflating the absence of semantics with participant-imposed organization. 
While a mutable canvas may better reflect real-world interfaces, we prioritize experimental control to better pose the canvas as our semantically not meaningful spatialization. We expect different design decisions for the canvas interface likely would change the results, thus the Canvas condition should not be interpreted as a validation of the semantic similarity metric itself. 

The Tree condition introduces a design constraint, as there is no canonical hierarchy for analysis steps. Rather than imposing an arbitrary structure, participants were given a flat list and a short three-minute structuring phase for lightweight grouping. This duration was chosen based on observations from pilot studies, where participants organized the list into two to three clusters of points within that time. This controlled structuring period ensured comparable organization effort across participants; however, we describe this as a limitation of our study design in Section~\ref{subsec:limitations}. 

\subsubsection{Datasets and Tasks}
We used three datasets (diamonds\cite{noauthor_diamonds_2025}, Airbnb listings\cite{noauthor_airbnb_2025}, and vehicle car sales\cite{noauthor_vehicle_2025}) which share similar semantics. Each has a mix of categorical and numeric columns centered around item-level pricing. For each condition, participants were provided an identical free-form EDA brief per condition: \textit{``Explore this dataset, surface interesting observations, and articulate promising next steps.''} All conditions were initialized with an identical, curated set of analets. Participants were permitted to use AI-assisted coding (OpenAI GPT-4.1~\cite{gpt41}) to reflect realistic, modern workflows.

Targeted goal-driven analysis often suggests an analytic solution. Open-ended EDA, however, does not enjoy this benefit; there is no immediate suggestion of a starting point nor of toolset. \meridian targets this scenario, and thus we employ a free-form EDA task rather than a goal-directed task in our user study.

\subsubsection{Procedure}
Each session comprised three blocks (one per condition). Before each block, participants received an orientation to common controls. Blocks lasted 25--40 minutes, depending on how quickly participants completed the post-analysis survey. Each block had at least 17 minutes of analysis and was run as a think-aloud. We captured screen/audio, researcher notes, interaction logs, and post-analysis ratings. Full protocol details, including CSO comparative review procedures, are provided in Appendix~\ref{app:interface-protocol}.

\subsubsection{Measures}
\label{sec:metrics-eval}
To evaluate user perceptions of \meridian, we developed a 20-item survey organized around ten domains. We adapted six domains from the Creativity Support Index (CSI)~\cite{cherry_quantifying_2014} (Immersion, Expressiveness, Exploration, Results Worth Effort, Enjoyment, and Collaboration), reframing them for the context of EDA. For example, ``exploration'' here refers to querying subsets, inspecting distributions, and testing hypotheses, rather than activities typical of generative domains like brainstorming visual forms. While the changes depart from CSI's original format, they align with survey design best practices that advocate for domain-grounded construct framing~\cite{kitchenham_preliminary_2002,boateng_best_2018}.
We supplemented these with four wayfinding-related domains: orientation, planning, recall of history, and global coverage. These were chosen because they roughly correspond to the types of navigational knowledge posed by critical challenges in navigating analytic spaces~\cite{siegel_development_1975, kim_acquisition_2021}. Each domain was prefaced with a brief definition clarifying our working interpretation.
Each domain comprised two Likert-scale items, yielding 20 questions total. Participants completed the survey after each condition, and after the final analysis, they were given a chance to review and revise their responses across all three conditions concurrently. Instrument validation details (expert review and item refinements) are provided in Appendix~\ref{app:interface-protocol}.

\subsubsection{Counterbalancing and Assignment}
We used a counterbalanced assignment to mitigate order and carryover effects across conditions and datasets. Full assignment rationale and schedules are provided in Appendix~\ref{app:interface-protocol}.

\subsection{Results}
We analyzed the study data using a combination of transcripts, video recordings, and structured survey responses. The first author conducted a reflexive thematic analysis to identify design-relevant insights. In parallel, we examined participant behavior through their responses on the subjective Likert-scale questions covering wayfinding and application support domains.

Our findings are organized into two parts: first, we present results and reflections for our comparative structured observation highlighting how participants evaluated different spatial encodings across wayfinding-relevant dimensions. Second, we report emergent themes from the qualitative data that cut across conditions and extend beyond the predefined constructs, offering additional design considerations and avenues for future exploration.

\begin{figure}[htbp]
    \centering  \includegraphics[width=\linewidth,keepaspectratio]{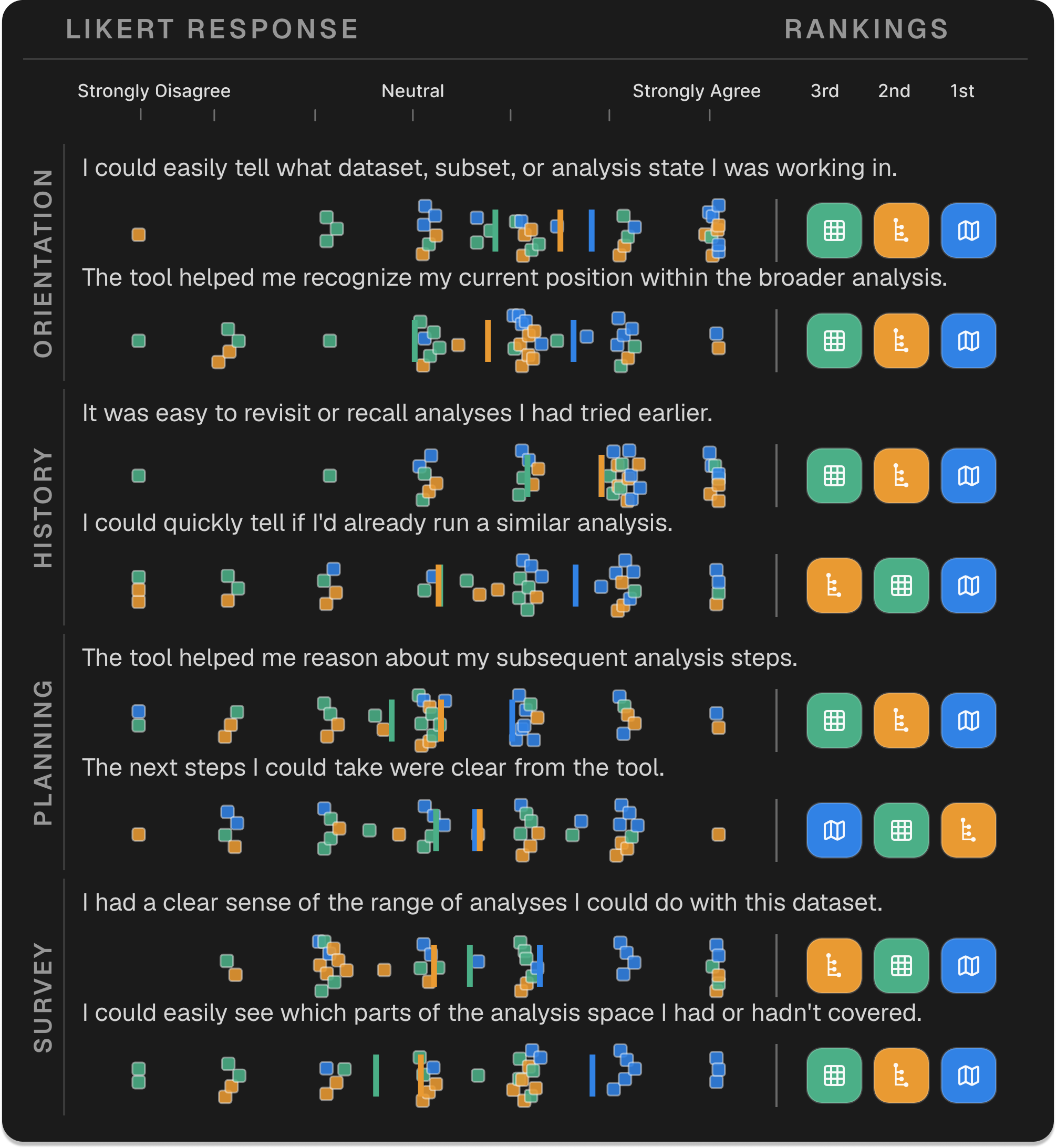}
    \caption{Subjective ratings and per-item rank preferences across interfaces. Left panels show beeswarm Likert responses (1=Strongly disagree, 7=Strongly agree) with interface means; right panels show rank preferences (\textit{1st}/\textit{2nd}/\textit{3rd}). Decimals were allowed during post-task reranking. The other domains are shown in the appendix \ref{app:interface-non-wayfinding}}
    \label{fig:interface-survey}
\end{figure}

We supplement our reflexive thematic analysis of think-aloud transcripts ($P1$--$P12$) with survey ratings (7-point Likert; $n{=}12$). Figure~\ref{fig:interface-survey} is the primary quantitative reference for wayfinding outcomes in this section. Detailed non-wayfinding items from the adapted CSI domains are summarized in Appendix~\ref{app:interface-non-wayfinding}.

\subsection{Wayfinding: Global Survey, Orientation, and Interpretability}

Participants strongly favored the Map condition for tasks involving global wayfinding (Figure~\ref{fig:interface-survey}). On a seven-point Likert scale, it received the highest average ratings for perceiving the range of possible analyses ($M=5.13$; 9/12 participants ranked it first) and for seeing covered versus uncovered regions of the analysis space ($M=5.66$; 12/12 ranked it first). 
Similarly, participants reported high confidence in using the Map to identify their current dataset, subset, or analysis state ($M=5.65$), and to situate themselves within the broader analytical process ($M=5.47$; 11/12 ranked it first). 
By contrast, the Tree condition offered competitive local orientation ($M=5.33$ for identifying current state), likely due to persistent textual labels, while the Static Canvas lagged across most wayfinding-related items ($M=3.47$--$4.67$). 

Qualitative interviews support this pattern. Participants consistently described the Map as a mental model for organizing their analytical journey. For example, $P1$ recounted the use of the Map to organize their analysis ``\textit{analyzing my fuel is here, analyzing my prices is here, analyzing my condition is here.}'' Others experienced the Map as a real-time progress tracker. $P4$ referred to it as a ``\textit{representation of what areas of the analysis has been explored},'' while $P8$ affirmed, ``\textit{I could easily see which parts of the analysis space have [been covered].''} However, the Map's interpretability was not universal. Some participants struggled to make sense of its structure. $P2$ found the layout ``\textit{hard to tell why things are apart}'' and noted that empty regions were not actionable ``\textit{because it's not interpretable \dots{} what exactly the distance metric is.}'' In contrast, the Tree offered immediate readability: $P8$ appreciated that, ``\textit{each item in the list had a textual description upfront \dots{} you could just see from there}'', and $P3$ emphasized the benefit of reduced interaction: ``\textit{the names of the tree nodes were visible. So it was much easier to see without doing a bunch of hovering.}'' $P1$ summarized the broader cognitive style divide: ``\textit{I am way better at graphs than I am with text, so lists are much, much worse for me than anything that's mappy.}''
\vspace{-1em}

\subsection{History and Similarity Recall}
The ability to revisit or recall earlier analyses was rated equally high for both the Map and Tree conditions ($M=5.75$ for both; Figure~\ref{fig:interface-survey}), suggesting that both global survey cues (Map) and textual summary (Tree) provide effective memory scaffolds. 
However, when it came to recognizing whether a similar analysis had already been run, the Map outperformed both alternatives ($M=5.49$), ahead of the Static Canvas ($M=4.12$) and the Tree ($M=4.10$). 
This supports the idea that \textit{spatial distance-as-similarity} helps surface near-duplicates during exploratory analysis. Qualitative responses reinforce this interpretation. Participants found that the Map's clustering naturally exposed structural overlap: $P3$ noted that similar analyses appeared ``\textit{basically right on top of each other},'' making redundancy obvious at a glance. Meanwhile, the Tree's labeled structure supported a different kind of memory recall, helping participants return to specific, named steps. As $P3$ explained, the Tree helped: ``\textit{just go back and find that thing by name.}''
\vspace{-1em}

\subsection{Naturally Imposed Reading Order Affects Planning}

Participants rated the Map best for supporting \textit{reasoning about subsequent steps} ($M=4.85$), while the Tree led narrowly in clarity of what to do next ($M=4.52$), with the Map close behind ($M=4.47$) and the Canvas lower ($M=4.08$; Figure~\ref{fig:interface-survey}). 
This pattern suggests a tradeoff between \textit{strategic planning} and \textit{immediate actionability}: the Map better supports reasoning about where to go in the analysis space, while the Tree provides a clearer order for what to do next. 

Both the Tree and Canvas provided strong ordinal cues that encouraged linear progression. Participants often approached them as a checklist: ``\textit{just pick the next analet}'' or, in $P8$'s words, ``\textit{press all the buttons!}'' $P4$ described creating child nodes that were a ``\textit{strict superset of independent variables compared to the parents},'' and using this structure to navigate top-down through related analyses. As they noted, ``\textit{The tree organization gives me a sense of strong relations between different analysis and how they're actually related.}'' 

By contrast, the Map required a two-step decision: first selecting a region of the space (a semantic cluster), then choosing an action within it. This decoupling led to more strategic exploration for some but introduced planning friction for others. $P4$ found the \textit{embedding cursor}, the Map's preview mechanism, opaque for planning (``\textit{I'm not sure what's the interaction signal that should be given \dots{} for this step I want to consider a different variable}.''), yet the survey values suggest that many participants were nevertheless able to translate this global structure into plausible next moves. 
For example, $P1$ found the cursor a helpful tool for next steps, using it to probe new questions within a region. 

\subsection{Behavioral Breadth vs.\ Subjective Wayfinding: A Counter-Intuitive Split}
Despite receiving the lowest subjective ratings on many wayfinding measures, the Canvas condition yielded the widest analytical breadth, as measured by convex-hull area over the semantic embedding (Averages of the Canvas $0.42$, Map $0.27$, Tree $0.24$, where $1.0$ is normalized to $P2$'s Canvas, the largest observed). $P9$'s results were excluded due to irregularities in the study setup. 

This surprising result points to what we term an enumerative to-do affordance: participants often described the Canvas as encouraging them to ``\textit{push all the buttons}'' or ``\textit{pop bubble wrap},'' quickly sampling a wide variety of analets, because the interface imposed minimal interpretive friction. The Tree interface produced a milder version of this effect; its serialized list structure invites users to work through items sequentially but its nested refinement encourages depth within local branches, limiting lateral movement. 

In contrast, Map participants frequently chose to deepen exploration within a single cluster before jumping to new regions. This behavior led to lower immediate hull area but supported more intentional local comparison and strategic reconnaissance. Several participants used the Map's layout to anchor their attention and assess coverage more deliberately.

We hypothesize that, over longer sessions, the Map's explicit cues for unvisited space and negative-space affordances would support broader dispersion. However, in short-form studies like ours, the interpretive cost of the Map may in fact promote analytical fixation, trading breadth for depth.
\vspace{-1em}

\subsection{Mental Models and Affordances}
A consistent theme across participants was that alignment between the interface and their mental model strongly influenced their effectiveness and satisfaction. For example, $P3$ described their analytic process as a hierarchical ``\textit{drill-down},'' beginning ``\textit{very, very broad}'' and narrowing to ``\textit{a number or a set of numbers}.'' The Tree interface resonated with this structure, providing a natural fit for their way of thinking. $P2$ similarly framed their process as progressive refinement, grouping related perspectives: ``\textit{a histogram or a scatterplot... different views of the same data}'', and creating ``\textit{children}'' that ``\textit{incorporate more information}.''

Participants who identified as mind-mappers sought both hierarchy and interval association. As $P6$ put it: ``\textit{The whole exploration is a mind map}.'' For many, spatial metaphors shaped their interface preferences: $P1$, for instance, preferred ``\textit{graphs}'' over text-based structures. Yet interpretability also mattered deeply. $P2$ wanted semantic control over the layout, expressing a desire to ``\textit{click and drag something to be further away, and that would in turn adjust the weights}.'' 

Other participants blended spatial and variable-centric mental models. For example, $P11$ described constructing an internal representation in which ``\textit{I mentally draw each column as a node. And then I draw a graph, asking how connected are these two. I then want to group them based on the question I'm trying to ask}.'' She later clarified that this takes the form of a bipartite graph between variables and analysis questions, where edges encode the variables involved in a given question. This structure closely mirrors the Semantic Map, suggesting that the Map condition aligns with how some participants already conceptualize their analytical space.

Importantly, some participants wanted more than navigational support; they wanted to author and shape the space itself. While \meridian~does support the integration of new notebook cells into the Semantic Map, this feature was disabled during the study to focus participant engagement on navigational behaviors. Several participants noted this limitation and expressed interest in more interactive and generative capabilities that would allow them to actively construct, not just traverse, their analytic landscapes.

\vspace{-0.75em}
\subsection{Additional Cross-Cutting Findings}
\pheading{Trust and algorithm aversion.} $P5$, $P6$, and $P12$ noted that encountering buggy or illogical code eroded their trust in the interface. As $P6$ stated: ``\textit{Can I trust it? That’s my question}.'' This breakdown in confidence shifted their behavior from open-ended exploration to cautious verification, and in some cases, even avoidance. $P5$, for instance, opted to switch to a different notebook entirely rather than engage in debugging.


\pheading{Abstraction and integrated outputs.} For some participants, code served merely as a means to an end rather than the primary focus. As $P5$ put it, ``\textit{the code feels like not that important to me \dots{} I wanna operate at a higher level of abstraction}.'' Embedding visual previews and analysis results directly within the Map, while still allowing drill-down access to code when needed, would better align with this preferred, abstraction-oriented workflow.

\pheading{Meaningful spatial semantics.} Spatial layout must convey interpretable structure. $P5$ described the Map as ``\textit{a thousand x better}'' than the list, citing how proximity intuitively signaled similarity. However, the absence of labeled regions made navigation difficult. Several participants requested regional labels or legends to help anchor their understanding and guide exploration more effectively.


\pheading{What counts as analysis.} $P8$ drew a clear distinction: analysis is the interpretation of outputs, not the act of generating code. They suggested that for the Map to feel like a true analysis tool rather than merely a navigation aid, it should surface meaningful artifacts such as charts, statistics, or AI summaries of findings---not merely underlying actions.

\pheading{A simple grid can outshine a smart map.} $P1$ found the Canvas more expressive than the Map in practice, citing two reasons: (i) its simpler, seemingly more fundamental analets made compositional logic easier to grasp; and (ii) lacking any semantic commitments, the grid ``\textit{is just the grid},'' which avoided mismatched expectations when the meaning of spatial axes was unclear. 
\section{Discussion}
\subsection{Trade-offs between spatial representations}
\subsubsection{Balancing Playful Exploration with Disciplined Reasoning}
Across conditions, participants found \meridian's semantic map engaging, but also more cognitively demanding than a tree or grid. The map excelled at revealing global structure (supporting orientation and survey tasks) and made near-duplicate analyses visually salient via distance-as-similarity. However, several participants hesitated when converting that overview into an immediate next step. In contrast, the tree and grid imposed a natural ``reading order'': top-to-bottom or left-to-right, which functioned as a to-do list. This structure reduced decision overhead and made next steps feel obvious, even though it offered fewer signals about global coverage or conceptual relationships. Together, these findings reveal a core design trade-off: spatial encodings that prioritize \textit{survey} afford strategic reasoning (e.g., tracking coverage, promoting diversity), while those with strong \textit{route} affordances lower planning friction but can encourage shallow, enumerative interactions.


\subsubsection{Designing for Deliberate Action}
Our most counter-intuitive result is empirical: the \textit{Static Canvas} produced the largest behavioral breadth (as measured by convex hull area), despite scoring lowest on subjective wayfinding. Participants described the grid as ``\textit{push all the buttons}'' where breadth emerged quickly from enumerative sampling. The tree condition exhibited a milder version of the same effect: its sequential structure created a natural to-do flow, encouraging participants to move linearly down a branch. 

By contrast, the map prompted more deliberate action. Participants often explored one conceptual neighborhood in depth before jumping elsewhere, supporting local comparisons and strategic reconnaissance, but reducing short-term dispersion. We caution against interpreting the grid's breadth advantage as evidence against semantic maps. Rather, this highlights an early-stage \textit{decision-cost} issue: global awareness demands cognitive effort unless paired with low-friction next steps.

A more productive design goal is to blend the map's survey power with optional, lightweight \textit{routes} that reduce decision overhead. For example, ``tour this cluster'' to serialize immediate neighbors, ``contrast nearest neighbors'' to encourage micro-comparisons, or ``jump to farthest relevant region'' to diversify perspective.

\subsubsection{Manipulating the Space as a First‑class Analytic Act}
Participants repeatedly expressed a desire to actively \textit{shape} the analytic space by dragging items to assert or refute similarity, weighting dimensions of relevance, or pinning anchors to restabilize the surrounding layout. We draw inspiration from prior work in semantic interaction, where spatializations serve as manipulable hypothesis surfaces that adapt as the model is steered~\cite{endert_semantic_2015}. This approach opens several promising directions for research and design:

\begin{itemize}
  \item \textbf{Metric steering.} Allow users to influence the similarity metric by demonstration, dragging points closer or farther apart—to personalize what counts as ``nearby'' within the \cellql~embedding.
\item \textbf{Region-level operators.} User feedback in the study indicates that labeling clusters would help users interpret the semantic meaning of different regions in the map. Treating these neighborhoods as composable, first-class sets could also enable a richer space of interaction techniques. For example, users might ``\textit{summarize this cluster},'' ``\textit{derive a median-split boundary},'' or ``\textit{diff these two groups}.'' Such operators could build on ideas from semantic interaction~\cite{endert_semantic_2012}, allowing users to manipulate regions directly while expressing higher-level analytical intent. 
  \item \textbf{Hybrid lattice representations.} Provide an optional, density-aware snapping that converts the 2D map into a more stable lattice structure. This preserves local proximity while supporting a lightweight, spatialized reading order for planning and scanning.
  \item \textbf{Explicit dimensions.} Drawing inspiration from systems like \cite{suh_luminate_2024} which make design spaces manipulable through discrete valued axes, \meridian~could improve interpretability and steerability by making these axes explicit. 
\end{itemize}

\subsection{Design Implications}

While \meridian~is designed for notebook environments, our findings reveal broader design patterns relevant to any system where users generate analytic actions. EDA tools should aim to expose both ordinal and interval structure to support wayfinding across analysis spaces. At the same time, designers should recognize when simplicity, such as a tree or grid, is the right affordance, especially early in a session when low decision overhead matters most.

More broadly, we advocate designing for EDA as navigation within a conceptual space, drawing inspiration from diverge–converge cycles from the design process~\cite{peng_simply_2018}. Evaluations could reflect this framing by separating dimensions like enjoyment, flow, exploration, and wayfinding using inventories tailored to analysis. Lastly, we see manipulability of the analytic space itself as a powerful substrate for mixed-initiative systems, enabling users and agents to collaborate through shared spatial meaning.
\vspace{-0.5em}
\subsection{Limitations}
\label{subsec:limitations}

Our interface study investigates how three representations shape navigation through a fixed set of candidate analets. Across conditions, participants could preview and materialize matched candidate analyses, while Bearings and live insertion of newly authored cells were disabled. The results therefore characterize the navigational affordances of the Semantic Map, Tree, and Static Canvas rather than the end-to-end history construction, recommendation quality, or the effectiveness of Bearings.

\rr{Given our \textit{n}=12 participant count, we lack the power necessary to make any causal claims, but we find the qualitative results of the Comparative Structured Observation (CSO) to be useful and valid.} Using a CSO across multiple interfaces imposes natural constraints on study design. While sessions averaged \textit{2h 12m} with a single mid-session break, and the generous stipend (\$75) and break helped mitigate fatigue, prolonged exposure to a novel tool may still have affected later-session engagement. Each interface condition lasted approximately \textit{17 minutes}, consistent with prior EDA studies~\cite{zgraggen_investigating_2018,battle_characterizing_2019}, but this relatively short window emphasizes early-stage behaviors (e.g., the enumerative sampling seen in Canvas) and may under-represent long-horizon planning and refinement.

While our evaluation demonstrates that \cellql~can faithfully capture data semantics for most typical operations, its declarative design trades the full expressivity of Python for normalization, restricting its ability to represent code like machine learning model fitting. The fixed framing set (100 seed analets) is similarly a pragmatic compromise: it cannot represent the full combinatorial EDA space and may under-represent some user-specific intents depending on the sampling strategy. Additionally, generative features such as Fixes depend on local seed density, meaning recommendations may lose coherence in sparse regions where the system is forced to extrapolate beyond the convex hull of seeded examples.


To reduce cold-start friction and ensure comparable idea availability across conditions, we pre-seeded each interface with the same set of \textit{analets} and permitted AI coding assistance. However, buggy or illogical outputs occasionally shifted participants' focus from exploration to verification, diverting cognitive effort. 

Given these constraints, and limitations such as the fixed period for tree layout, our emphasis is on thematic findings. We interpret quantitative metrics like breadth and depth cautiously, recognizing that short blocks and code artifacts can confound behavioral dispersion. A longitudinal, in-the-wild deployment would allow us to examine how analysts construct and evolve their own analytic spaces from scratch, an opportunity several participants (e.g., $P1$, $P6$) explicitly requested, expressing interest in ``\textit{building the map from the ground up}.''
\vspace{-0.5em}
\subsection{Future Work}
\rr{Our work opens new possibilities for designing spatial representations that mirror how analysts reason during EDA. While \meridian 
demonstrates the benefits of structured, semantically grounded spatial layouts, several promising directions emerged from participant feedback and limitations of the present system. Below, we outline opportunities to broaden \meridian's representational expressiveness, adaptivity, and alignment with user cognition.}

\rr{
\pheading{Richer spatial encodings.} Participants wanted to represent many-to-one rationales: multiple steps supporting a single conclusion, or collections that function as conceptual groups. Future work could extend the map with n-ary links (e.g., hyperedges connecting several analets to one knowledge node) and set-level constructs like group badges. These would support cross-cutting analyses while preserving the cognitive split between route and survey reasoning.
}


\rr{
\pheading{Composable analets.} Participants consistently wanted to start with simple building blocks and compose them into more complex steps. Future systems could surface the parts of an analet (selection, transforms, aggregation, comparison) as modular, pickable components, something like an ``analysis color picker" that lets analysts remix pieces from different analets. Factoring \cellql~primitives (\texttt{REPEAT}/\texttt{BRANCH}/\texttt{CONCAT}) into a structured palette with copy-with-binding and copy-as-value semantics would enable rapid construction while still preserving provenance.
}

\pheading{Longitudinal studies.} Free-form EDA naturally takes longer than what we can capture in a proctored lab session. A longitudinal deployment would let us see how tool preferences shift across analysis phases (focus, backtracking, goal-switching) and how analysts reshape the mapped space to suit their evolving needs.


\vspace{-0.5em}
\section{Conclusion}
EDA is not just about generating visualizations, but about \textit{navigating} a conceptual space of ideas, hypotheses, and interpretations. Yet most tools externalize results without exposing the structure of the analytic journey itself. Our work introduces \meridian, a system that models EDA as navigation through an analysis space. By representing analytic steps in a structured form and embedding them in a similarity-based layout, \meridian~makes landmark, route, and survey knowledge visible and actionable. This enables analysts to reason not only about individual outputs, but also about neighborhoods, coverage, and trajectory over time. 
Our findings highlight a fundamental design tension between survey-oriented spatial encodings, which support strategic reasoning, and route-oriented encodings, which reduce planning friction. Rather than privileging one over the other, we argue for interfaces that support fluid movement between expansive exploration and deliberate focus. More broadly, we advocate treating analytic space itself as a manipulable substrate. Making the structure of analysis visible and steerable opens new opportunities for adaptive recommendation, semantic interaction, and collaborative sensemaking. We hope this work encourages future systems to move beyond chart-by-chart iteration toward trajectory-aware, spatially grounded inquiry.

\appendix
\section{Appendix A: CellQL Formalism and Preliminary Evaluation}
\label{app:cellql-formalism}

\noindent
This appendix provides additional detail on the internal representation used by \meridian. We briefly describe the \cellql~formalism, report a preliminary analysis of its behavior under syntactic variation, and outline how the system constructs the seeded analysis space used for layout stability and guidance.

\subsection{Formal Language: \cellql}
\label{subsec:cellql}

Each analet is associated with a \cellql~representation, which provides a unified, syntax-invariant specification of the cell's data consumption. The primary design goal of \cellql~is to isolate the structure of the data view required to produce an output: the columns accessed, filters applied, and transformations performed.

\cellql~is designed to concisely represent structural patterns common in EDA queries. It operates as a lightweight orchestration layer that coordinates generation of standard SQL queries. We model \cellql~formally as:

\[
\textit{CellQL} ::= \textit{unit} \mid \textit{REPEAT}(\cdot) \mid \textit{BRANCH}(\cdot) \mid \textit{CONCAT}(\cdot) \quad 
\]
\[
\textit{unit} := \textit{SQLSpec}
\]
The language relies on four compositional primitives:

\begin{enumerate}
    \item \textbf{Unit Template:} The leaf nodes of a \cellql~specification are standard SQL \texttt{SELECT} statements containing handlebars-style placeholders (e.g., \texttt{\{\{col\}\}}). While we employ helper functions like \texttt{BIN\_COUNT1D} to capture visualization intents (like histograms), these compile directly into standard SQL aggregation and binning operations.
    
    \item \textbf{REPEAT:} An explicit looping operator that emits queries for every item in a list (e.g., columns or variable pairs). This operator captures the analyst's intent to iterate over data dimensions, distinguishing a systematic scan of variables from a series of unrelated queries.
    
    \item \textbf{BRANCH:} A compile-time conditional that selects specific query templates based on schema facts (e.g., \texttt{TYPE(col) = NUMERIC}). Unlike a runtime SQL \texttt{CASE} statement which alters values within a row, \texttt{BRANCH} determines the structural shape of the query itself, for example, choosing between a binning query for numeric data and a value count for categorical data.
    
    \item \textbf{CONCAT:} A grouping operator that bundles independent queries into a single execution block representing parallel views or dashboards where distinct logical units (e.g., a chart and a summary table) are presented side-by-side.
\end{enumerate}

We considered several alternatives before arriving at this design. Raw \textbf{Python code} is sensitive to surface-level variation. A stacked bar chart implemented in Matplotlib and Altair share little syntax, yet perform identical analytic work. Embeddings over Python source can scatter semantically equivalent steps, undermining map coherence.

Grounding similarity in \textbf{visual form} presents the opposite problem. A stacked bar chart and a pie chart look different, but both group by a categorical variable, aggregate counts, and normalize to proportions. Wu et al.\ term this shared layer \textit{design-specific transformations} \cite{wu_design-specific_2024}. Embeddings over chart-type labels can miss this analytic equivalence and become sensitive to superficial styling changes.

Raw \textbf{SQL} is closer, but struggles with repeated structures common in EDA. A scatterplot matrix \cite{noauthor_scatterplot_nodate} is one conceptual operation (``compare every column pair''), yet compiles to many \texttt{SELECT} statements joined by \texttt{UNION ALL}. Embeddings over this SQL can be dominated by repeated boilerplate and query length rather than the underlying analytic pattern. Similar ``iterate over columns'' behavior appears in common operations like \texttt{df.describe()}, \texttt{df.info()}, and \texttt{df.hist()}.

\cellql~addresses this through combinators that preserve template structure. Rather than dozens of queries, a scatterplot matrix is expressed as \texttt{REPEAT} over column pairs with shared templates. The embedding sees the strategy directly ("apply this template across pairs"), capturing analyst intent rather than execution verbosity.

\subsection{\cellql Preliminary Evaluation}

An ideal representation should capture the semantics of an analysis. If two items are placed close together in the map, that proximity should reflect true semantic similarity. Two qualities are critical for such a representation: consistency and differentiation. Consistency means that surface-level syntactic changes do not alter meaning. For example, a scatterplot written in Altair should be treated the same as a scatterplot written in Matplotlib. Differentiation means that functional changes are reflected as greater distances such that a histogram and a scatterplot of the same variables, or a correlation matrix versus a descriptive summary, are represented as distinct.

In order to test whether \cellql~satisfies these criteria, we normalize Python notebook cells into \cellql, embed them alongside two comparison modalities (raw Python code and English natural-language paraphrases), and then evaluate the resulting embeddings along two core dimensions: (i) \textit{semantic consistency}: does \cellql~preserve meaning across syntactic variants, remaining invariant to superficial code differences? and (ii) \textit{semantic differentiation}: does the representation separate analyses according to their functional intent, such that embeddings reflect meaningful variation in analytic purpose?

\subsubsection{\cellql Evaluation Pipeline}
\begin{figure*}[t]
    \centering
    \includegraphics[width=\textwidth]{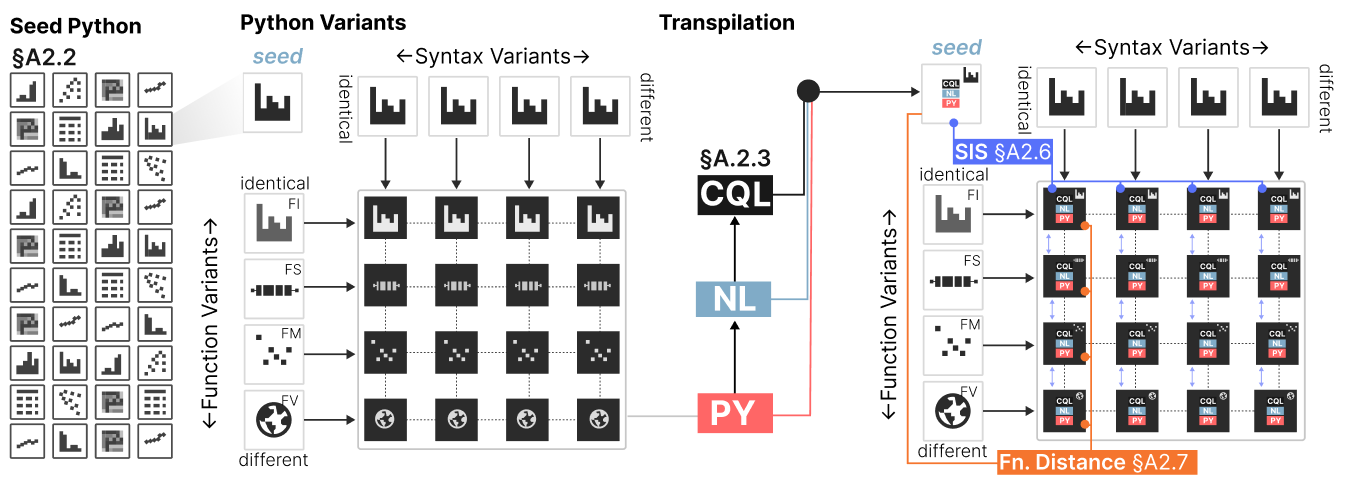}
    \caption{\textbf{Variant generation and evaluation pipeline for \cellql.} Starting from seed Python analets (left), we generate controlled \textit{syntax variants} (columns) and \textit{function variants} (rows: FI, FS, FM, FV). Each variant is transpiled into Natural Language and \cellql (center), producing three parallel modalities. These are embedded into a shared semantic space, where we measure (i) \textbf{syntax invariance} (SIS, §A.2.6), capturing whether functionally identical variants remain close, and (ii) \textbf{functional distance} (§A.2.7), capturing whether distance increases with semantic change. The resulting structure tests whether \cellql preserves semantic identity while separating distinct analyses.}
    \label{fig:variant_gen}
\end{figure*}

To evaluate whether \cellql~produces representations that are consistent under refactoring and differentiating under functional change, we implement an end-to-end evaluation pipeline (Figure~\ref{fig:variant_gen}). The pipeline transforms raw notebook cells into normalized representations, generates controlled variants, and compares their embeddings across three modalities (Python code, \cellql, and natural language descriptions). Each stage leaves artifacts for reproducibility (JSON bundles, logs, outputs), and variant band assignments are validated using an \textit{LLM-as-a-Judge}~\cite{zheng_judging_2023} to reduce noise from execution errors or syntactic quirks. We describe each stage below.

\subsubsection{Generating Variant Analets}
\label{subsec:variants}

We first create a diverse corpus of analytic steps. An LLM produces $N{=}100$ realistic notebook cells covering common EDA tasks (e.g., \texttt{df.info}, histograms, scatterplots, group-by aggregations, transforms). \rr{The authors reviewed this initial list of generated analets for quality and diversity regenerating 2 analets which did not meet proper EDA functionality. These 100 analets represent the ``seed cells" which form a baseline \textit{analet}.} 

To test consistency and differentiation, we generate 12 \textit{variants} per seed, spanning four functional bands (FI, FS, FM, FV). Together with the unchanged baseline, this yields 13 items per seed. As shown in Figure~\ref{fig:variant_gen}, syntax variation is organized across columns while functional variation is organized across rows. Bands are defined as follows: 

\begin{itemize}[leftmargin=1em]
  \item \textbf{FI} (pure refactor): renaming/reordering/chaining; \textit{identical} output (same rows/columns/dtypes/order; same plot data/mapping).
  \item \textbf{FS} (small, deterministic): schema preserved and tiny controlled changes (e.g., \texttt{head(200)}, fixed‑seed \texttt{sample}, numeric rounding, stable sort; plot style tweaks).
  \item \textbf{FM} (moderate, related): derived features, within‑group transforms, added encodings; meaning shifts but remains clearly related; modality may change.
  \item \textbf{FV} (structural/task shift): pivot/aggregate/melt/correlation; density/hexbin vs.\ scatter; modality typically changes.
\end{itemize}

Variant candidates are proposed by Claude Sonnet~4 \cite{noauthor_claude_2025} and judged by Gemini 2.5-Pro \cite{noauthor_gemini_2025}, which evaluates whether each variant meets the intended band criteria. Prompts disallow “magic” helpers and prefer minimal, self-contained code. This dual-model process was used to ensure robust generation and acceptance.

\subsubsection{Transpilation to \cellql}
\label{subsec:transpile}

Each accepted variant is transpiled into two additional modalities: a concise natural-language description and a normalized \cellql~program. As illustrated in the center panel of Figure~\ref{fig:variant_gen}, this yields three parallel representations for each item: Python, natural language, and \cellql. The NL summary scaffolds the structured transpilation, following grammar-prompting principles~\cite{wang_grammar_2023}. These generations allow us to compare how well different representations preserve semantic meaning of similar variants.

To assess the fidelity of \cellql~translations, we employed an LLM-as-judge approach using three independent models (GPT-5-nano, Gemini 2.5 Flash Lite, and Claude 4.5 Haiku) as evaluators. Each judge was prompted to determine whether the generated \cellql~query captured the data semantics of the original Python code: whether it operated on approximately the same columns, used similar aggregations, performed comparable derived calculations, and applied equivalent row filters.  A translation was marked as passing if a majority of judges (2 of 3) deemed it semantically equivalent. Across 1300 Python-\cellql~pairs the translation pipeline achieved an overall pass rate of 86.6\%, demonstrating that \cellql~can reliably capture the data access patterns and requirements of typical data analysis code.

The cases where translation failed were rarely random errors, but rather instances where the procedural nature of the Python code clashed with the declarative constraints of \cellql. This disconnect manifests most sharply in statistical inference and machine learning, where Python performs complex model fitting (e.g., training a Random Forest) or hypothesis testing (e.g., ANOVA p-values) that transcend \cellql's simple algebraic aggregations. Similarly, certain visualization nuances are often lost like KDE parameters or granular row-level renderings (e.g., rugplots) are not captured. Finally, failures arise in structural and metadata operations, where Python inspects memory usage or reshapes dimensionality (e.g., stack/unstack)—tasks that are fundamentally distinct from querying data values. We view these failures as opportunities to improve \cellql~perhaps through a more extensive function library like the Hypothesis Grammar~\cite{suh_grammar_2023} or through a better prompted transpilation. However, we still find good transpilation performance through many traditional EDA analets.


\subsubsection{Embeddings and Modalities}
\label{subsec:modalities}

All three modalities are embedded into the same semantic space using OpenAI’s \texttt{text-embedding 3-small} (1536-D)~\cite{noauthor_openai_2025}. As shown in the right panel of Figure~\ref{fig:variant_gen}, these aligned embeddings allow us to compare how each representation organizes syntax-preserving and function-changing variants. Using a single model across modalities avoids same-model bias between generation and embedding. Cosine distance is used for measure computation; UMAP~\cite{mcinnes_umap_2020} is applied only for visualization.


\subsubsection{Results}
\label{subsec:metrics}
Let $d_m(\cdot,\cdot)$ be cosine distance in modality $m\!\in\!\{\text{Python},\text{CellQL},\text{Natural Language}\}$.

\subsubsection{Semantic consistency (syntax invariance)}
\label{subsec:results-sis}

\[
\mathrm{SIS}
= 1 - \frac{d(\text{seed},\,\mathrm{FI\ bucket})}
{\min\{\,d(\text{seed},\,\mathrm{FS}),\ d(\text{seed},\,\mathrm{FM}),\ d(\text{seed},\,\mathrm{FV})\,\}}
\]\vspace{0.25em}

Syntactic invariance (SIS) evaluates whether \textit{functionally identical} (FI) variants remain close to their seed despite surface-level refactors. Figure~\ref{fig:variant_gen} highlights this comparison in blue: SIS asks whether items that differ syntactically but not functionally remain tightly clustered around the seed. For each seed and modality, SIS compares the mean embedding distance from the seed to FI variants (characterized by syntax buckets: Identical/Similar/Med/Far) against the nearest mean distance to any functionally different band (FS/FM/FV). The resulting score lies in $[0,1]$, where higher values indicate stronger syntactic invariance (i.e., functionally identical variants remain closer to the seed than any functionally different variants). Our SIS metric is adapted from \textit{behavioral invariance} testing in model evaluation~\cite{ribeiro_beyond_2020}, which probes systems with \textit{label-preserving} edits (e.g., synonym substitutions, formatting changes, or neutral distractors) and expects them to maintain consistent outputs. 

We report mean SIS by bucket, accompanied by 95\% percentile bootstrap confidence intervals over seeds (10{,}000 resamples). Because SIS is \textit{bounded} in $[0,1]$ and often \textit{heavy–tailed} (with mass near 0 under our hard normalization), bootstrap CIs are preferable to $t$–intervals and better reflect uncertainty of the mean. Intervals for \cellql~do not overlap with those for Natural Language or Python in any bucket, indicating clear separation at the mean level, and suggesting that \cellql~better preserves semantic identity under syntactic variation.

\begin{table}[h]
\centering
\small
\begin{tabular}{lccc}
\toprule
 & \textbf{Similar} (95\% CI) & \textbf{Med} (95\% CI) & \textbf{Far} (95\% CI) \\
\midrule
\cellql~(SQL)  
& \shortstack{\textbf{0.399}\\{\scriptsize [0.316,\;0.483]}} 
& \shortstack{\textbf{0.417}\\{\scriptsize [0.333,\;0.499]}} 
& \shortstack{\textbf{0.395}\\{\scriptsize [0.311,\;0.484]}} \\

Natural Language   
& \shortstack{0.243\\{\scriptsize [0.185,\;0.302]}} 
& \shortstack{0.223\\{\scriptsize [0.164,\;0.285]}} 
& \shortstack{0.227\\{\scriptsize [0.167,\;0.289]}} \\

Python        
& \shortstack{0.093\\{\scriptsize [0.052,\;0.141]}} 
& \shortstack{0.098\\{\scriptsize [0.055,\;0.148]}} 
& \shortstack{0.087\\{\scriptsize [0.045,\;0.133]}} \\
\bottomrule
\end{tabular}
\caption{Mean SIS by syntax bucket with 95\% percentile bootstrap CIs over seeds (10k resamples). Higher is better. Lower SIS indicates that syntax is distorting semantics.}
\label{tab:sis-mean-ci}
\end{table}

\textbf{\cellql~is most invariant to syntax changes.} Across buckets, \cellql~ ($\approx$\,0.395–0.417) is consistently higher than Natural Language ($\approx$\,0.223–0.243) and Python ($\approx$\,0.087–0.098). In practical terms, pure refactors move FI points far less in \cellql space than in NL or code spaces. Large surface changes degrade NL \& code more. Even at \textit{Far}, \cellql remains $\approx$\,0.395 while Natural Language and Python drop to $\approx$\,0.227 and $\approx$\,0.087, respectively. The Med/Similar ordering is close and not perfectly monotonic for some modalities (e.g., \cellql: 0.417\,>\,0.399; Natural Language: 0.223\,<\,0.243). This is expected because (i) generation-time \texttt{level} is a proxy for surface distance and can be noisy, and (ii) SIS normalizes by the \textit{nearest} function-different band per seed, which varies across seeds. The overall modality ranking (\cellql $>$ Natural Language $>$ Python) holds under median aggregation and fixed-band denominators (reported in the supplement).

\subsubsection{Functional distance (ordered separation)}
\label{subsec:functional-distance}

To evaluate whether distances from the seed's baseline (F0) increase with \textit{functional} change, we compute cosine distances between F0 and each of its band variants: FI (identical), FS (slight), FM (moderate), FV (very different). Figure~\ref{fig:variant_gen} highlights this ordering in orange, corresponding to increasing functional distance away from the seed. We then summarize across seeds.\footnote{All distances are computed in the original 1536-D embedding space; 2D UMAP is for visualization only. Lower is closer to F0.}
\begin{table}[h]
  \centering
  \small
  \begin{tabular}{lcccc}
  \toprule
   & \textbf{FI} & \textbf{FS} & \textbf{FM} & \textbf{FV} \\
  \midrule
  \cellql (SQL)
  & \shortstack{\textbf{0.099}\\{\scriptsize [0.077, 0.125]}}
  & \shortstack{\textbf{0.127}\\{\scriptsize [0.103, 0.153]}}
  & \shortstack{\textbf{0.255}\\{\scriptsize [0.232, 0.280]}}
  & \shortstack{0.313\\{\scriptsize [0.289, 0.337]}} \\

  Natural Language
  & \shortstack{0.079\\{\scriptsize [0.071, 0.087]}}
  & \shortstack{0.083\\{\scriptsize [0.074, 0.093]}}
  & \shortstack{0.235\\{\scriptsize [0.216, 0.254]}}
  & \shortstack{\textbf{0.315}\\{\scriptsize [0.290, 0.339]}} \\

  Python
  & \shortstack{0.087\\{\scriptsize [0.076, 0.098]}}
  & \shortstack{0.036\\{\scriptsize [0.027, 0.048]}}
  & \shortstack{0.166\\{\scriptsize [0.145, 0.188]}}
  & \shortstack{0.289\\{\scriptsize [0.262, 0.317]}} \\
  \bottomrule
  \end{tabular}
  \caption{Mean cosine distance from F0 to each functional band. For each
  modality--band combination, 95\% percentile-bootstrap confidence
  intervals for the mean were computed by resampling seeds with replacement
  10{,}000 times and taking the 2.5th and 97.5th percentiles of the
  bootstrapped means. Bands represent increasing intended functional change
  from FI to FV.}
  \label{tab:functional-mean-ci}
  \end{table}

Distances \textit{increase} with functional change in \cellql~and Natural Language, with the largest separation at FV (task or structure shifts). In contrast, Python exhibits:  \textit{FS $<$ FI}, consistent with our SIS results,  indicating that code embeddings are more influenced by refactoring-style changes than by slight functional differences.  \cellql~better preserves ``same function'' neighborhoods while still differentiating truly different analyses.

\subsubsection{Limitations of the \cellql Evaluation}
This validation is intentionally preliminary and scoped. It uses LLM-generated seed cells and controlled variants rather than organically evolved notebooks, and semantic equivalence is assessed with multiple LLM judges rather than human annotation. The goal is not complete transpilation of arbitrary Python notebooks, but to test whether CellQL is locally adequate for common EDA operations and robust to surface-level refactoring. The 86.6\% pass rate should therefore be interpreted as evidence of useful adequacy within this scope, not full coverage. Failures cluster in procedural/model-centric code, metadata operations, and certain visualization details, which we treat as out-of-scope opportunities for future extension. 

\begin{figure*}[!t]
  \centering
  \includegraphics[
    width=\textwidth,
    keepaspectratio
  ]{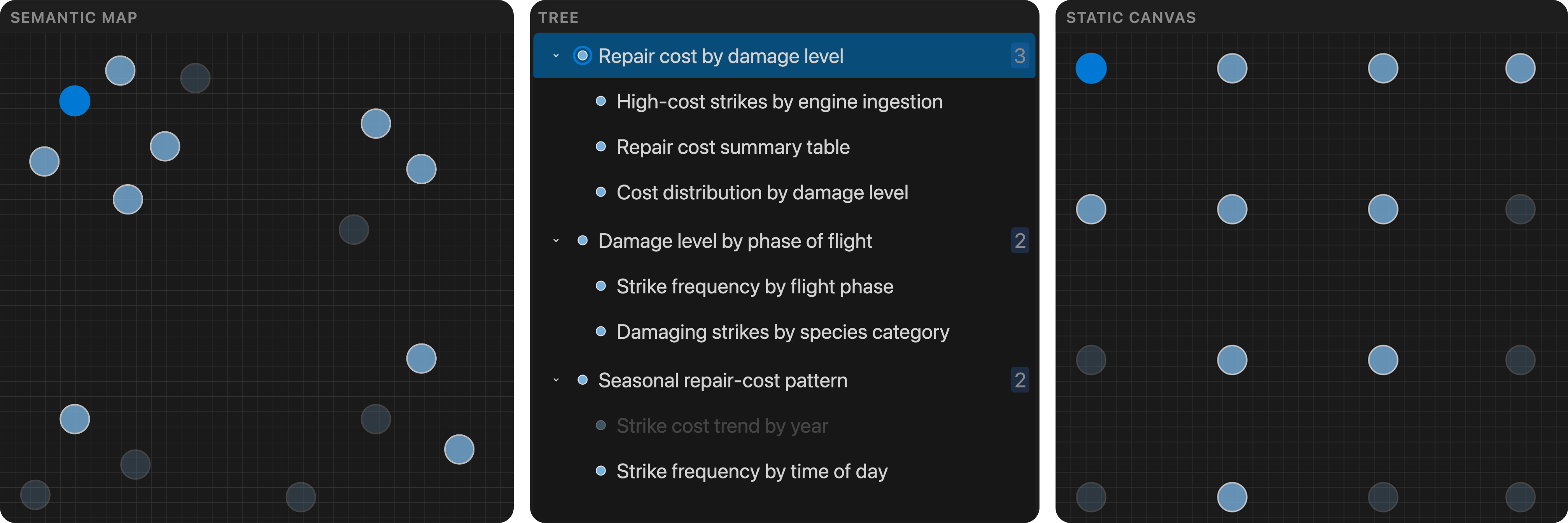}
  \caption{Side-by-side overview of the three interface conditions used in
  the study. The Semantic Map uses local proximity to represent similarity
  among analets; the Tree presents analets as a labeled, user-organized
  hierarchy; and the Static Canvas places analets in a fixed grid whose
  positions have no semantic meaning. The selected analet is shown in bright
  blue across all three conditions.}
  \Description{Three dark interface panels shown side by side. The Semantic
  Map positions circular analets according to semantic similarity. The Tree
  organizes labeled analets into expandable hierarchical groups. The Static
  Canvas places the same analets in a regular grid. A selected analet is
  highlighted in bright blue in each panel.}
  \label{fig:interface-conditions}
\end{figure*}

\subsection{Seed Set Construction}
\label{appsubsec:framing}
\noindent
Having defined the representation, we now describe how \meridian~instantiates an initial seed set for map stabilization and neighborhood guidance.

To seed the map with example analyses, \meridian~initializes each session with a curated set of approximately 100 seeded analets. This serves two purposes: (1) positioning user-authored analets within a stable semantic reference frame, and (2) supporting interaction techniques such as fixes (Section~\ref{subsec:navigation}) that probe underexplored regions using nearby examples.

\meridian~constructs this initial set using a recommendation process that combines schema information, lightweight statistical profiling, any available data dictionary, and analyst task framing. Following Jupybara's emphasis on pragmatic relevance \cite{wang_jupybara_2025}, \meridian~uses an LLM to generate a small set of stakeholder personas (e.g., policy analyst, data journalist) and goals (e.g., "compare urban and rural outcomes"), then derives corresponding seed analyses.

The goal of this seeding process is not exhaustive coverage of the combinatorially large \cellql~space, but sufficient span for stable positioning. In practice, plausible EDA operations are constrained by schema (column count, type mix, and meaningful groupings). We observed that layouts became reasonably stable around 40 diverse seeds, while larger sets improved the utility of neighborhood-dependent features such as Fixes. For the study, we used 100 seeds as a balance between layout quality, feature utility, and computational tractability. These framing analets do not directly populate the user-visible Map or Tree; they provide the reference scaffold for positioning and navigation.

\section{Appendix B: Extended Interface Study Protocol}
\label{app:interface-protocol}

\noindent
This appendix consolidates the protocol and supplemental quantitative details that are referenced in the main Interface Evaluation section. We present the session procedure first, then instrument development, assignment strategy, and finally non-wayfinding survey outcomes from the adapted CSI domains.

\subsection{Session Procedure}
Each session comprised three blocks (one per condition). Before each block, participants received a short orientation to common controls. Blocks lasted 25--40 minutes, depending on how quickly participants completed the survey. Each block included at least 17 minutes of analysis. All blocks were run according to the think-aloud protocol. We captured screen/audio, researcher field notes, and interaction logs. After each block, participants completed a brief survey. At the end of the session, we conducted a comparative review in which ratings were shown side-by-side; participants could adjust scores while explaining changes and pointing to concrete interface elements. This comparative review operationalized the CSO requirement for structured cross-condition reflection.

\subsection{Interface Condition Overview}
\label{app:interface-conditions}

Figure~\ref{fig:interface-conditions} provides a side-by-side view of the
three interface conditions. All conditions exposed the same candidate
analets and supported the same preview and materialization interactions,
but organized those analets differently. The Semantic Map positioned
analets according to local semantic similarity; the Tree exposed labeled,
user-organized hierarchical and ordinal relationships; and the Static
Canvas placed analets in a fixed grid whose spatial positions carried no
semantic meaning. The comparison therefore examines how these different
organizational structures shape navigation while holding candidate
availability constant.


\subsection{Survey Instrument Development}
To validate the survey instrument, we conducted a structured expert review with six participants with backgrounds in user research and data-analysis-tool design. Experts evaluated each item for relevance to its construct, wording clarity, and coverage completeness. To reduce anchoring bias, we clarified that construct definitions were design rationale rather than fixed semantics. Feedback prompted phrasing revisions for five of the twenty items before deployment.

\subsection{Counterbalancing and Assignment}

To mitigate order and carryover effects, we used a counterbalanced assignment strategy. For the three interface conditions (Map, Canvas, Tree), we employed a Williams (balanced Latin square) design with six orders: \texttt{ABC, ACB, BAC, BCA, CAB, CBA}. With $N{=}12$ participants, each order was assigned twice, ensuring that each condition appeared equally often in each ordinal position (four times in positions 1, 2, and 3).

Datasets were assigned using a balanced permutation schedule over three datasets (Vehicles, Diamonds, Listings) such that: (i) each dataset appeared exactly four times in each session position, and (ii) condition--dataset pairings were distributed as evenly as possible across the full $3{\times}3$ matrix. This ensured that no condition was systematically paired with a particular dataset or position.

The resulting assignment (Supplementary Material \textit{Study Procedure Diagram}) balances both condition order and dataset exposure across participants. This design reduces confounds from learning, fatigue, and ordering effects while preserving sensitivity to differences between interface conditions.

\subsection{Supplementary Interface Survey Results (Non-Wayfinding)}
\label{app:interface-non-wayfinding}
This subsection reports adapted CSI domains that were collected but are not central to the paper's wayfinding argument (Exploration, Expressiveness, Immersion, Enjoyment, Results Worth Effort, and Collaboration). Main-text quantitative interpretation focuses on wayfinding domains; this figure preserves completeness of reporting.

\begin{figure}[htbp]
  \centering
      \includegraphics[width=\linewidth,height=0.9\textheight,keepaspectratio]{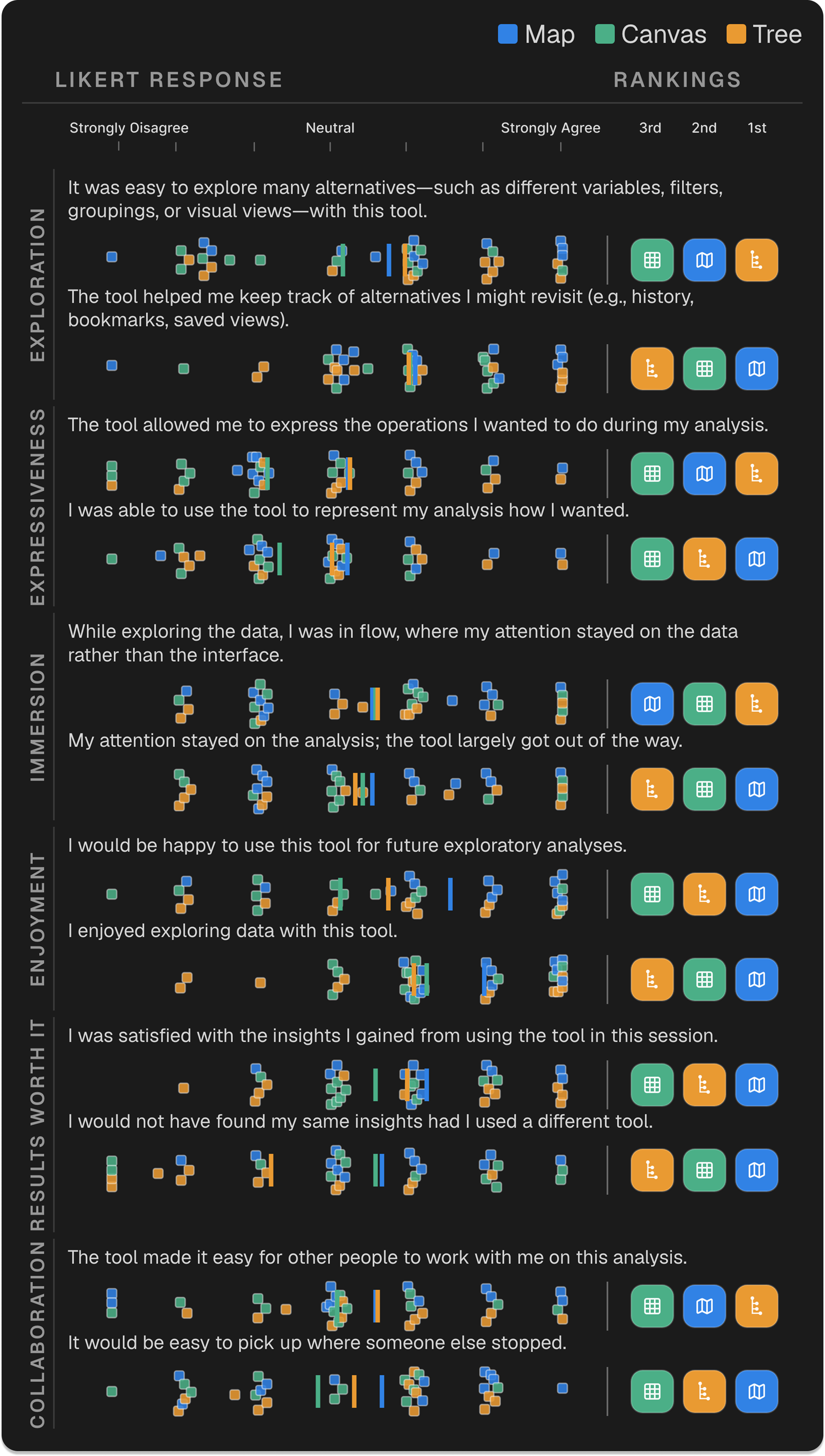}
      
  \caption{Supplementary adapted CSI outcomes (non-wayfinding domains). Main-text wayfinding claims are reported in Figure~\ref{fig:interface-survey}.}
  \Description{A dark-background dot plot compares participant responses
for the Semantic Map in blue, Static Canvas in green, and Tree in orange
across twelve survey items. The items cover six adapted Creativity
Support Index domains: exploration, expressiveness, immersion, enjoyment,
results worth effort, and collaboration. Each square represents one
participant's seven-point Likert response, and vertical colored marks
show condition averages. Icons on the right show the conditions' relative
rankings for each item. The Semantic Map ranks first on eight of the
twelve items, including keeping track of alternatives, representing an
analysis as desired, enjoyment, satisfaction with insights, and resuming
another person's work. The Tree ranks first on the other four items,
including exploring many alternatives, expressing intended operations,
flow, and collaboration.}
  \label{fig:interface-survey-csi}
\end{figure}

\clearpage

\bibliographystyle{ACM-Reference-Format}
\bibliography{references, references-2}
\end{document}